\documentclass[
prd,
reprint,
nofootinbib,
superscriptaddress,
 amsmath,
 amssymb,
 aps,
floatfix,
showkeys
]{revtex4-2} 

\makeatletter
\def\@fnsymbol#1{\ensuremath{\ifcase#1\or *\or \dagger\or \ddagger\or
   \mathsection\or \mathparagraph\or \|\or **\or \dagger\dagger
   \or \ddagger\ddagger \or *** \else\@ctrerr\fi}}
\makeatother

\usepackage[colorlinks=true,citecolor=blue,filecolor=blue,linkcolor=blue,urlcolor=blue]{hyperref}
\usepackage{graphicx}
\usepackage{tabularx}
\usepackage{dcolumn}
\usepackage{bm}
\usepackage{xcolor}
\usepackage{float}
\usepackage{adjustbox}
\usepackage{lineno}
\usepackage{array}
\usepackage{siunitx}
\usepackage{enumerate}
\usepackage{multirow}
\usepackage{url}
\usepackage{upgreek}
\usepackage{physics}
\usepackage{enumitem}
\newcommand{\tsinghuaphy}{\affiliation{Department of Physics \& Center for High Energy Physics, Tsinghua University, Beijing 100084, China}}

\newcommand{\tsinghuaephy}{\affiliation{Key Laboratory of Particle and Radiation Imaging (Ministry of Education) \& Department of Engineering Physics, Tsinghua University, Beijing 100084, China}}

\newcommand{\ustc}{\affiliation{State Key Laboratory of Particle Detection and Electronics, University of Science and Technology of China, Hefei 230026, China}}

\newcommand{\ustcimp}{\affiliation{
Department of Modern Physics, University of Science and Technology of China, Hefei 230026, China}}

\newcommand{\westlake}{\affiliation{School of Science, Westlake University, Hangzhou 310030, China}}

\newcommand{\sysuphy}{\affiliation{School of Physics, Sun Yat-Sen University, Guangzhou 510275, China}}

\newcommand{\sysuifcen}{\affiliation{Sino-French Institute of Nuclear Engineering and Technology, Sun Yat-Sen University, Zhuhai 519082, China}}

\newcommand{\cuhk}{\affiliation{School of Science and Engineering, The Chinese University of Hong Kong (Shenzhen), Shenzhen, Guangdong, 518172, P.R. China}}

\newcommand{\buaasp}{\affiliation{School of Physics, Beihang University, Beijing 100083, China}}

\newcommand{\buaalab}{\affiliation{Beijing Key Laboratory of Advanced Nuclear Materials and Physics, Beihang University, Beijing 100191, China}}

\newcommand{\sanmen}{\affiliation{CNNC Sanmen Nuclear Power Company, Zhejiang 317112, China}}

\newcommand{\cevns}{CE$\nu$NS }
\newcolumntype{C}[1]{>{\centering\arraybackslash}p{#1}}

\begin{document}


\title{Reactor neutrino liquid xenon coherent elastic scattering experiment}

\author{Chang Cai}\tsinghuaphy

\author{Guocai Chen}\sanmen

\author{Jiangyu Chen}\sysuifcen

\author{Rundong Fang}\buaasp

\author{Fei Gao}
\email[Corresponding author: ]{feigao@tsinghua.edu.cn}
\tsinghuaphy

\author{Xiaoran Guo}\ustc\ustcimp

\author{Jiheng Guo}\buaasp

\author{Tingyi He}\altaffiliation[Also at ] {College of Physics, Chongqing University, Chongqing 401331, China}\cuhk

\author{Chengjie Jia}\altaffiliation[Now at ] {Department of Physics, Stanford University, Stanford, CA 94305, USA}\tsinghuaphy

\author{Gaojun Jin}\sanmen

\author{Yipin Jing}\altaffiliation[Also at ] {School of Physics, Jilin University, Jilin 130012, China}\tsinghuaphy

\author{Gaojun Ju}\sanmen

\author{Yang Lei}\tsinghuaphy

\author{Jiayi Li}\tsinghuaphy

\author{Kaihang Li}\tsinghuaphy

\author{Meng Li}\sanmen

\author{Minhua Li}\sanmen

\author{Shengchao Li}\westlake

\author{Siyin Li}\westlake

\author{Tao Li}\sanmen

\author{Qing Lin}\ustc\ustcimp

\author{Jiajun Liu}\sysuphy

\author{Minghao Liu}\tsinghuaphy

\author{Sheng Lv}\sanmen

\author{Guang Luo}\sysuphy

\author{Jian Ma}\tsinghuaphy

\author{Chuanping Shen}\sanmen

\author{Mingzhuo Song}\tsinghuaphy

\author{Lijun Tong}\ustc\ustcimp

\author{Xiaoyu Wang}\westlake

\author{Wei Wang}\sysuifcen\sysuphy

\author{Xiaoping Wang}\buaasp\buaalab

\author{Zihu Wang}\sanmen

\author{Yuehuan Wei} \email[Corresponding author: ]{weiyh29@mail.sysu.edu.cn} \sysuifcen

\author{Liming Weng}\sanmen

\author{Xiang Xiao}\sysuphy

\author{Lingfeng Xie}\tsinghuaphy

\author{Dacheng Xu}\altaffiliation[Now at ]{Physics Department, Columbia University, New York, NY 10027, USA.}\tsinghuaphy

\author{Jijun Yang}\westlake

\author{Litao Yang}\tsinghuaephy

\author{Long Yang}\sanmen

\author{Jingqiang Ye}\cuhk

\author{Jiachen Yu}\ustc\ustcimp

\author{Qian Yue}\tsinghuaephy

\author{Yuyong Yue}\sysuifcen

\author{Bingwei Zhang}\sanmen

\author{Shuhao Zhang}\tsinghuaphy

\author{Yifei Zhao}\tsinghuaphy

\author{Chenhui Zhu}\ustcimp

\collaboration{RELICS Collaboration}
\noaffiliation

\date{\today}

\begin{abstract}
Coherent elastic neutrino-nucleus scattering (CE$\nu$NS) provides a unique probe for neutrino properties Beyond the Standard Model (BSM) physics. REactor neutrino LIquid xenon Coherent Scattering experiment (RELICS), a proposed reactor neutrino program using liquid xenon time projection chamber (LXeTPC) technology, aims to investigate the \cevns process of antineutrinos off xenon atomic nuclei. In this work, the design of the experiment is studied and optimized based on Monte Carlo (MC) simulations. To achieve a sufficiently low energy threshold for \cevns detection, an ionization-only analysis channel is adopted for RELICS. A high emission rate of delayed electrons after a big ionization signal is the major background, leading to an analysis threshold of 120\,photo-electrons in the \cevns search. The second largest background, nuclear recoils induced by cosmic-ray neutrons, is suppressed via a passive water shield. The physics potential of RELICS is explored with a 32\,$\rm kg \cdot yr$ exposure at a baseline of 25\,m from a reactor core with a 3\,GW thermal power. In an energy range of 120 to 300\,PE, corresponding to an average nuclear recoil from 0.63 to 1.36\,keV considering the liquid xenon response and detector-related effect, we expect 4639.7 \cevns and 1687.8 background events. The sensitivity of RELICS to the weak mixing angle is investigated at a low momentum transfer. Our study shows that RELICS can further improve the constraints on the non-standard neutrino interaction (NSI) compared to the current best results. 

\begin{description}
\item[PACS numbers]
\item[Keywords]
\end{description}
\end{abstract}

\pacs{
    95.35.+d, 
    14.80.Ly, 
    29.40.-n,  
    95.55.Vj
}

\keywords{CE$\nu$NS, reactor neutrino, liquid xenon, LXeTPC, MC}

 \maketitle

\section{Introduction}
\label{sec:intro}

Coherent elastic neutrino-nucleus scattering (CE$\nu$NS) is a Standard Model (SM) weak neutral current process postulated in 1974\,\citep{Freedman:1973yd}. The incoming neutrino sees the nucleus as a whole during scattering with a cross-section proportional to the nucleon number squared in the full coherency regime when the neutrino energy is smaller than $\sim$10\,MeV. Despite this process having a relatively large cross-section, the detection is challenging due to the ultra-low deposited energy in keV or sub-keV nuclear recoil scale. This process had evaded experimental discoveries for decades until 2017 when it was first observed by the COHERENT collaboration with neutrinos produced by the Spallation Neutron Source (SNS) at Oak Ridge National Laboratory scattering off a CsI scintillator\,\citep{COHERENT:2017ipa}, and later confirmed with a liquid argon detector\,\citep{COHERENT:2020iec} in 2020. These measurements of \cevns offer new probes for diverse physical phenomena. 

\cevns can provide new constraints on the SM physics and the physics Beyond it\,(BSM). The weak mixing angle is related to the \cevns cross-section~\citep{Canas:2018rng}. Any deviation from the SM prediction in measurement can be an indication of new physics. \cevns is also a valuable tool to constrain nonstandard neutrino interactions (NSI)~\citep{Barranco:2007tz}, to search for sterile neutrinos~\citep{Anderson:2012pn}, to verify the existence of non-trivial neutrino electromagnetic properties~\citep{Dodd:1991ni}, to study the nuclear structure~\citep{Patton:2012jr} and understand the core-collapse supernova explosions~\citep{Suliga:2021hek}. In addition, the \cevns signal from the solar or atmospheric neutrinos is expected to be an irreducible background for the next-generation dark matter direct detection experiments, such as PandaX-xT~\citep{PandaX:2024oxq} and XLZD~\citep{Aalbers:2022dzr} using xenon as detection target\,\citep{XENON:2024ijk,PandaX:2024muv}. A precise measurement of this process, especially one with a xenon target, can provide valuable information for dark matter experiments digging into the so-called ``neutrino floor/fog"~\citep{Billard:2013qya, OHare:2021utq,Tang:2024prl}.

The observation of \cevns by the COHERENT collaboration opened the way to \cevns experimental studies. While the neutrino beam has been the primary source for the \cevns discovery, the anti-neutrino from nuclear reactors provides an alternative chance for \cevns detection. Therefore, the nuclear reactors with large anti-neutrino fluxes delivered with typical energies of a few MeV, which is in the full coherency regime of CE$\nu$NS, have been proposed as attractive complementary facilities for several \cevns experiments. Due to the considerably low recoil energy and complex background, the reactor \cevns is harder to detect. Many efforts are underway with reactor neutrinos, such as \citep{Colaresi:2022obx,CONUS:2020skt,CONNIE:2021ggh,RED-100:2019rpf,Ni:2021mwa,MINER:2016igy,NUCLEUS:2019igx,nGeN:2022uje,Karmakar:2024rpq,Fernandez-Moroni:2021nap,Ricochet:2021rjo,Bernard:2022zyf}. Benefiting from the higher flux and lower energy of reactor neutrinos compared with the SNS neutrino beam, the detection of reactor \cevns can provide a complementary sensitivity in most fields of the physical phenomena mentioned above through new couplings~\citep{Lindner:2016wff,Papoulias:2019xaw}, and also potential applications such as nuclear safeguards~\citep{vonRaesfeld:2021gxl}.

In this context, RELICS is a reactor neutrino experiment using liquid xenon time projection chamber (LXeTPC) technology. LXeTPC is one of the leading technologies for rare event searches, such as detecting dark matter and neutrinoless double beta decay~\citep{Aalbers:2022dzr, PandaX-4T:2021bab, XENON:2022ltv, LZ:2022lsv, nEXO:2021ujk, EXO-200:2019rkq}. Benefiting from its ultra-low threshold by using ionization-only channel\,\citep{XENON:2019gfn,PandaX:2022xqx} and reduced background, the LXeTPC is an excellent approach for reactor \cevns detection. The RELICS experiment is proposed to be located near the Sanmen Nuclear Power Plant in Zhejiang province, China. At a $\sim$25\,m baseline to the $\sim$3\,GW reactor core, the average neutrino flux available for RELICS is $\sim$10$^{13}$cm$^{-2}$s$^{-1}$.
 
This work is organized as follows. The operation principle and the design of the detector as well as the shielding system, are described in detail in section\,\ref{sec:design}. The simulation framework is introduced in section\,\ref{sec:sim_framework}.  In section\,\ref{sec:signal}, we predict the expected \cevns event rates from the Sanmen nuclear reactor. Section\,\ref{sec:bkgs} discusses a dedicated simulation of background contributions. The physics potential of RELICS is provided in section\,\ref{sec:physics}. A summary is presented in section\,\ref{sec:summary}.

\section{experimental designs}
\label{sec:design}

\subsection{RELICS detector}
\label{subsec:tpc}
The LXeTPC technology consists in measuring the energy deposited from particle interactions by simultaneously detecting the prompt scintillation light (S1) and delayed ionization charge (S2)~\citep{XENON:2010xwm}. The S1 is produced by the direct excitation of xenon atoms and the recombination of electron-ion pairs in LXe. The S2 is proportional to the number of liberated electrons extracted into gas xenon (GXe), which are then amplified via electroluminescence. Both S1 and S2 are recorded by photosensors, usually photomultiplier tubes (PMTs), on top and bottom arrays of the TPC. This allows a three-dimensional (3D) vertex reconstruction with a sub-centimeter precision, and the S2/S1 ratio can be used to discriminate between nuclear recoil (NR) and electronic recoil (ER) for background reduction. Typically, the energy threshold of LXeTPC is limited by the S1 signal in the range of $\sim$(2-3) photo-electrons (PEs). The inherent S2 amplification provides an alternative approach to decrease the energy threshold by utilizing the ionization-only channel, usually called the ``S2-only" analysis. This approach can reduce the energy threshold to a few ionization electrons (equivalent NR energy below 1\,keV$\rm_{nr}$) by sacrificing the NR/ER discrimination and depth information, which, in the case of \cevns detection, can be largely compensated by the high signal-to-background ratio profited from the intense reactor neutrino flux.

\begin{figure}[htbp]
\centering 
\includegraphics[width=.42\textwidth]{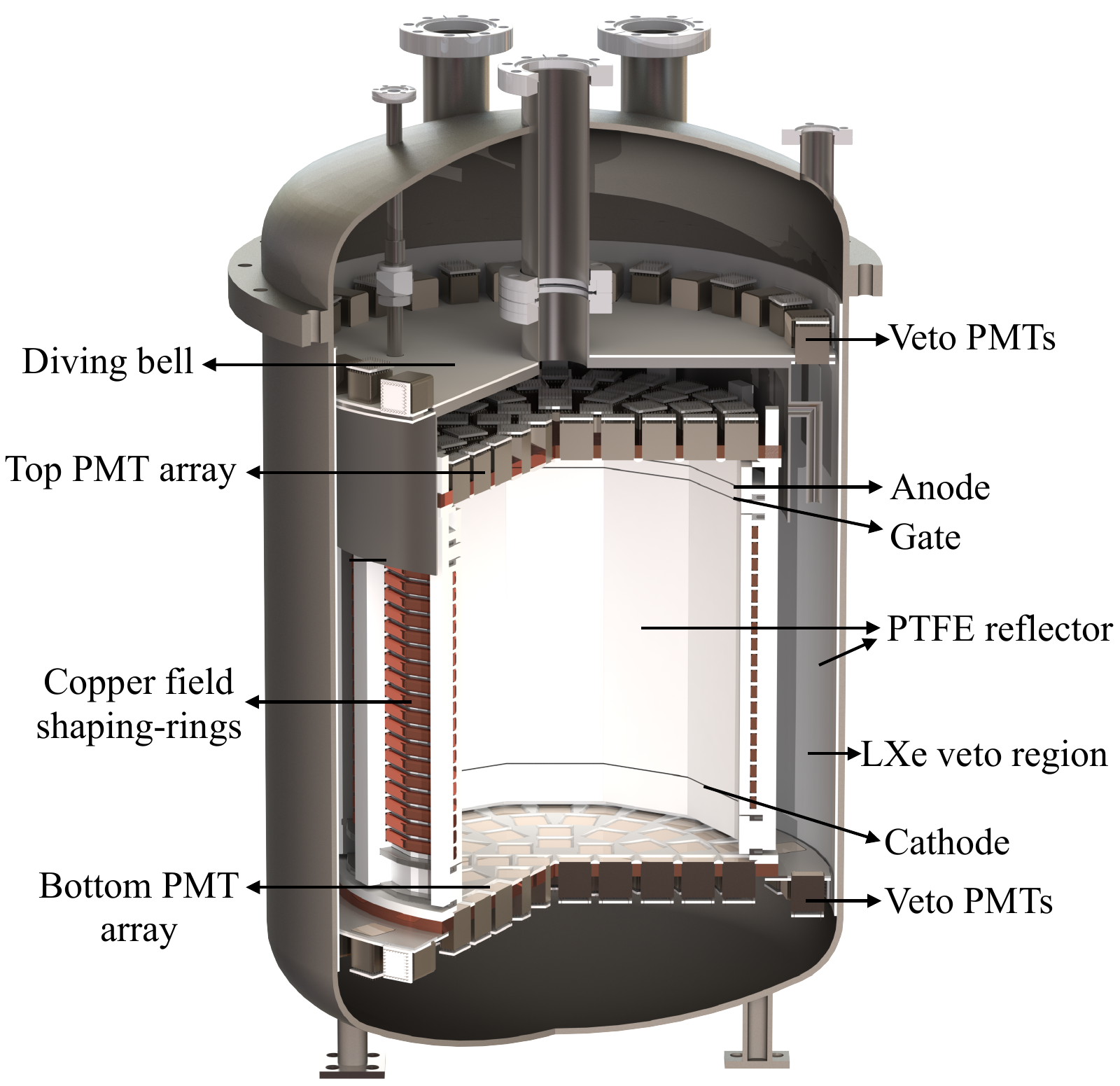}
\caption{The RELICS detector with the inner S.S. vessel, diving bell, veto PMTs, and the center TPC consisting of the top and bottom PMT arrays, anode, gate and cathode electrodes, copper field shaping-rings and PTFE reflector.}
\label{fig:tpc}
\end{figure}

A preliminary 3D mechanical design of the RELICS detector is shown in figure~\ref{fig:tpc}. A cylindrical TPC of 24\,cm in height and 28\,cm in diameter sits in a double-layer stainless steel (S.S) vessel, which provides a stable cryogenic environment. The target volume containing an active LXe of approximately 44\,kg is defined by 12 interlocking and light-tight PTFE (polytetrafluoroethylene, Teflon) panels. The active LXe is then viewed by two arrays of 64 Hamamatsu R8520-406 PMTs on top and bottom. A fiducial mass of 32\,kg LXe with only a radius selection of 12\,cm can be well-defined for \cevns search under ``S2-only" analysis. A ``diving bell" structure is mounted above the top PMT array to control the liquid level. Five etched S.S electrodes, including two grounding screenings on top and bottom to protect PMTs from high voltage (HV), anode, gate, and cathode to provide a nominal drift field of 500\,V/cm and an extraction field of 10\,kV/cm in GXe, are installed. To ensure the uniformity of the drift field, the target volume is encircled by several field-shaping rings made from oxygen-free high thermal conductivity (OFHC) copper. The TPC is surrounded by approximately 62\,kg of LXe, which acts as a 4$\pi$ active veto for background suppression. The LXe veto volume is instrumented with 48 R8520-406 PMTs. To achieve a high light collection, the inner surface of the inner vessel is covered with a thin PTFE foil. It is worth noting that the LXe veto is designed to reduce the background by $\sim$5 times by removing the coincidence events inside TPC and in the LXe veto layer. All materials for the detector construction will be carefully selected to ensure low intrinsic radioactivity.

\subsection{Shielding system}
\label{subsec:shield}

Near-reactor neutrino experiments usually operate with shallow overburdens. Thus, the high-energy cosmic-ray neutrons are supposed to be one of the main backgrounds for the NR searches from CE$\nu$NS. A sufficient neutron shielding structure, such as water, is required to slow down and absorb the neutrons. Meanwhile, the cosmic muon and its induced secondaries are expected to be another major background. A highly efficient active muon veto anti-coincidence system is also critical to the success of this type of experiment.

\begin{figure}
\centering
\includegraphics[width=0.48\textwidth]{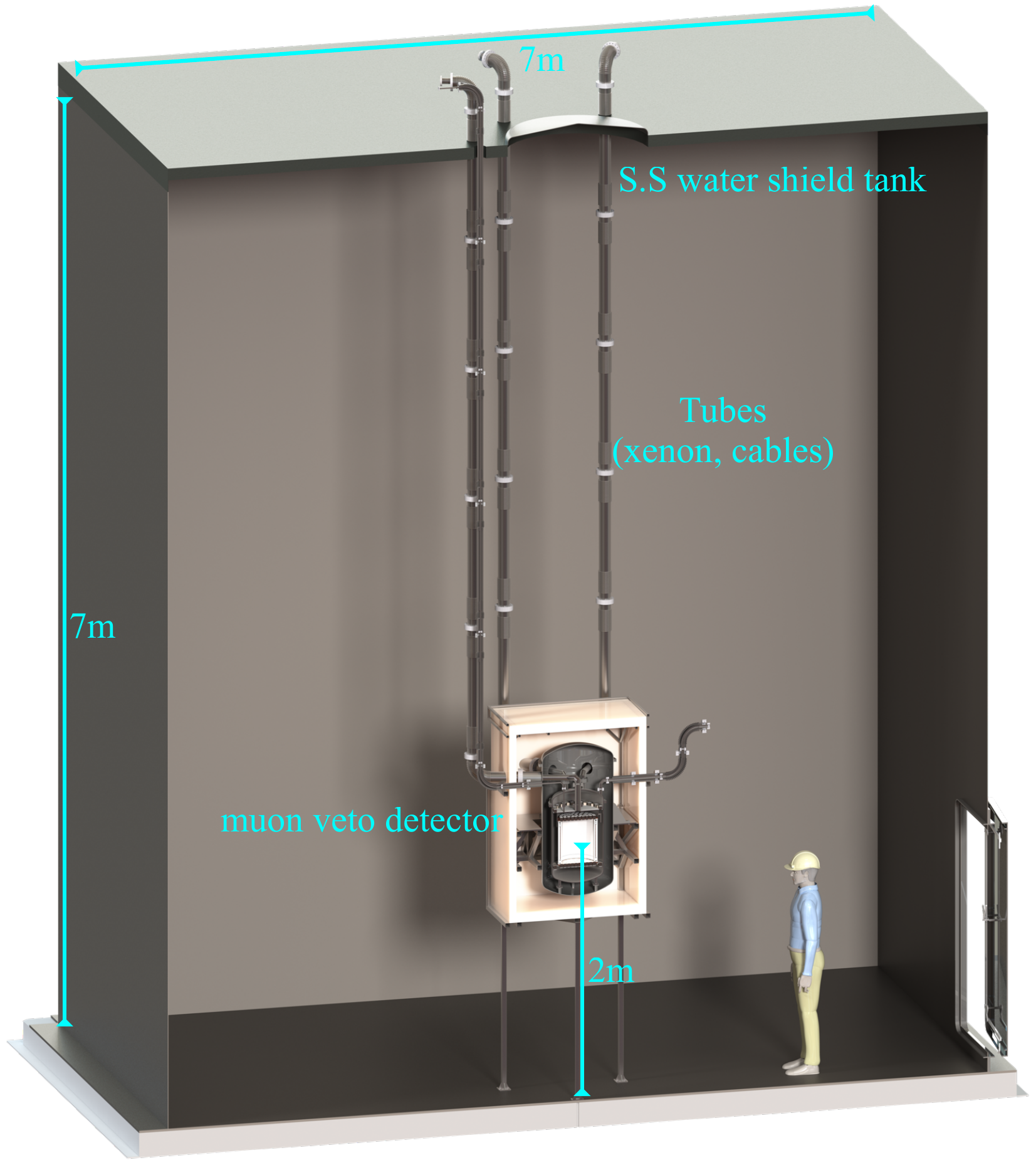}
\caption{The LXe detector with a 8\,cm thick PS muon veto detector inside a 7\,m$\times$7\,m$\times$7\,m S.S water shield tank. The center of the LXe detector is 2\,m above ground.}
\label{fig:shield}
\end{figure}

Figure~\ref{fig:shield} shows the design of the RELICS shielding system based on Monte Carlo (MC) simulations, which is detailed in section~\ref{sec:bkgs}. The design has been optimized to satisfy the requirement of the \cevns detection. 
The LXe detector is placed inside a shield of 7\,m$\times$7\,m$\times$7\,m of water contained in a S.S tank. A 5\,m water shield on top can suppress the cosmic-ray neutron-induced background to a controlled level. With such a water tank, the external neutrons and gammas from the laboratory environment and the nuclear reactor can be reduced to negligible levels.
In close to the LXe detector, a muon veto detector is installed. The veto detector is made of 8\,cm thick plastic scintillator (PS) equipped with wavelength-shifting fibers and silicon photomultipliers (SiPMs) to collect and detect the scintillation light. A high veto efficiency of 99\% is expected for the RELICS experiment.

\section{Simulation framework}
\label{sec:sim_framework}

The MC simulation framework of the RELICS experiment (\textsl{RelicsSim}) is built upon a Geant4-based BambooMC toolkit~\citep{Chen:2021asx}. Full detector geometry is implemented, including the detailed detector components and the shielding system. Various types of particle interactions are studied by simulation to evaluate the expected signal and background. Subsequently, \textsl{RelicsAPT}, which inherits from the \textsl{Appletree}~\citep{dacheng_xu_2023_8305639} framework developed by XENONnT collaboration~\citep{XENON:2024wpa}, is adopted to convert the energy depositions into observable S1 and S2 signals. The conversion is based on the intrinsic light yield (L$_y$) and charge yield (Q$_y$) in LXe, and convoluted with the detector-related signal detection and reconstruction effects. 

\subsection{Liquid xenon response}
\label{subsec:lxe_resp}

\begin{figure}[htbp]
\centering 
\includegraphics[width=.45\textwidth]{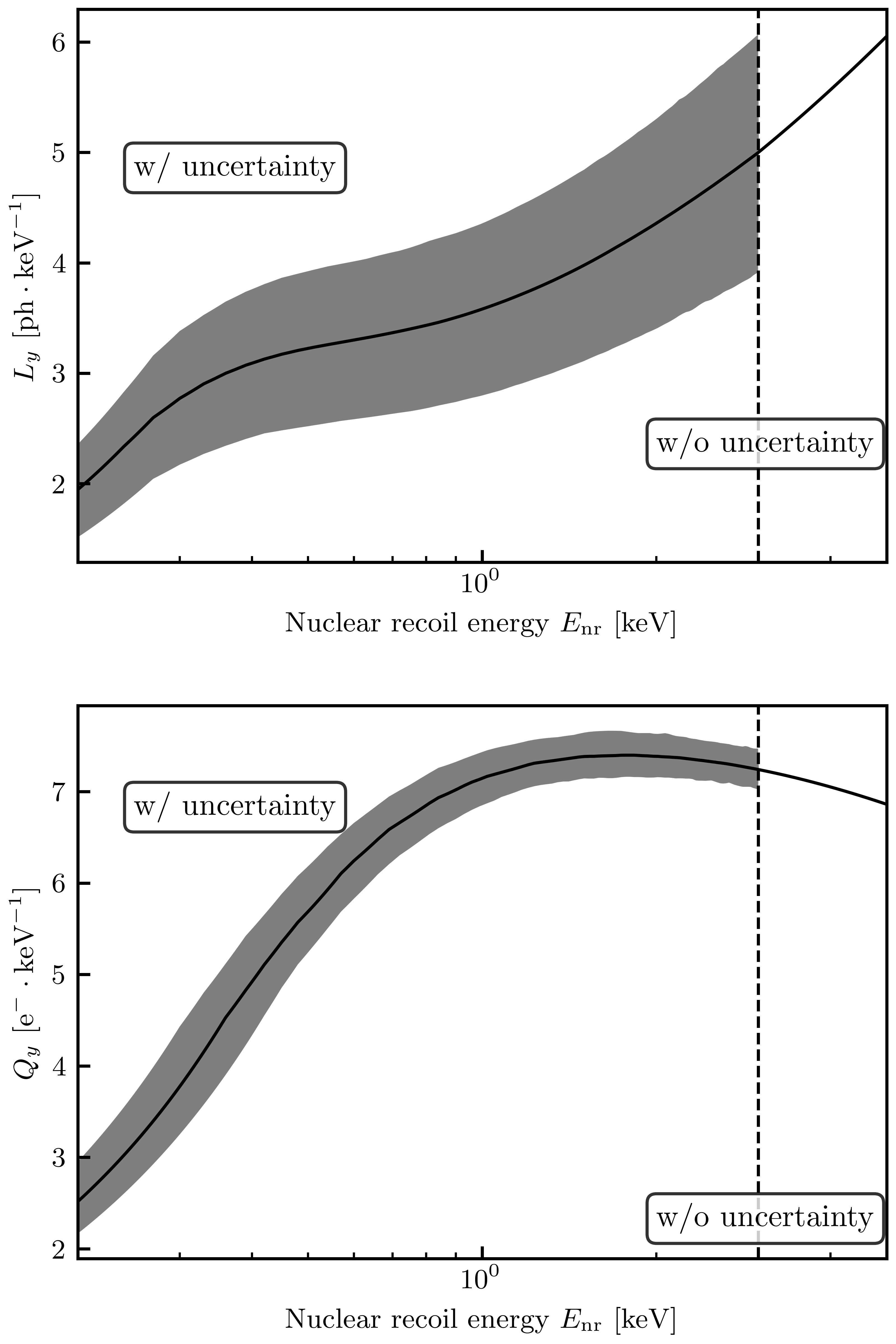}
\caption{The L$_y$ (upper panel) and Q$_y$ (lower panel) model of NR in \textsl{RelicsAPT}. For E$\rm_{nr}$ less than 3\,keV$\rm_{nr}$, the XENON1T yields model in $^{8}\rm B$ neutrino search\,\citep{XENON:2020gfr} is adopted, and scaled to NEST model at 3\,keV$\rm_{nr}$. For E$\rm_{nr}$ larger than 3\,keV$\rm_{nr}$, NEST model is used with 500\,V/cm drift field.}
\label{fig:lxe_response}
\end{figure}

Particle interaction in LXe will produce photons and electrons. The NEST model can characterize this process \,\citep{szydagis_review_2022} after considering the fluctuations in scintillation, ionization, electron-ion recombination, and their drift field dependence.  
In \textsl{RelicsAPT}, the L$_y$ and Q$_y$ from NEST\,(v2.3.6) at 500\,V/cm drift field are used with no uncertainty assigned for NRs larger than 3\,keV$\rm_{nr}$. To quantify the uncertainty of light and charge yields for \cevns events, the L$_y$ and Q$_y$ bands in XENON1T solar $^{8}\rm B$ neutrino search\,\citep{XENON:2020gfr} are adopted for NRs less than 3\,keV$\rm_{nr}$. L$_y$ and Q$_y$ bands are scaled to make the center curve connected to NEST at 3\,keV$\rm_{nr}$. The final L$_y$ and Q$_y$ of NRs applied in \textsl{RelicsAPT} are shown in figure~\ref{fig:lxe_response}. These light and charge yield responses and their uncertainties comparable to the XENON1T model are chosen as a benchmark model in this analysis to illustrate its impact on the sensitivity to \cevns and new physics beyond the Standard Model. For ERs, L$_y$ and Q$_y$ from NEST\,(v2.3.6) are used without uncertainties. In the 4$\pi$ LXe veto region, both L$_y$ and Q$_y$ are modeled with zero electric fields.


\subsection{Detector related effects}
\label{subsec:det_effect}

In addition to the intrinsic LXe response model, the detector-related effects must be considered when interpreting the observed S1 and S2 signals. The main detector parameters are considered based on the expected detector performance listed in Tab.\,\ref{tab:pars}, as a conservative estimation. 

\begin{figure}[htbp]
\centering 
\includegraphics[width=.45\textwidth]{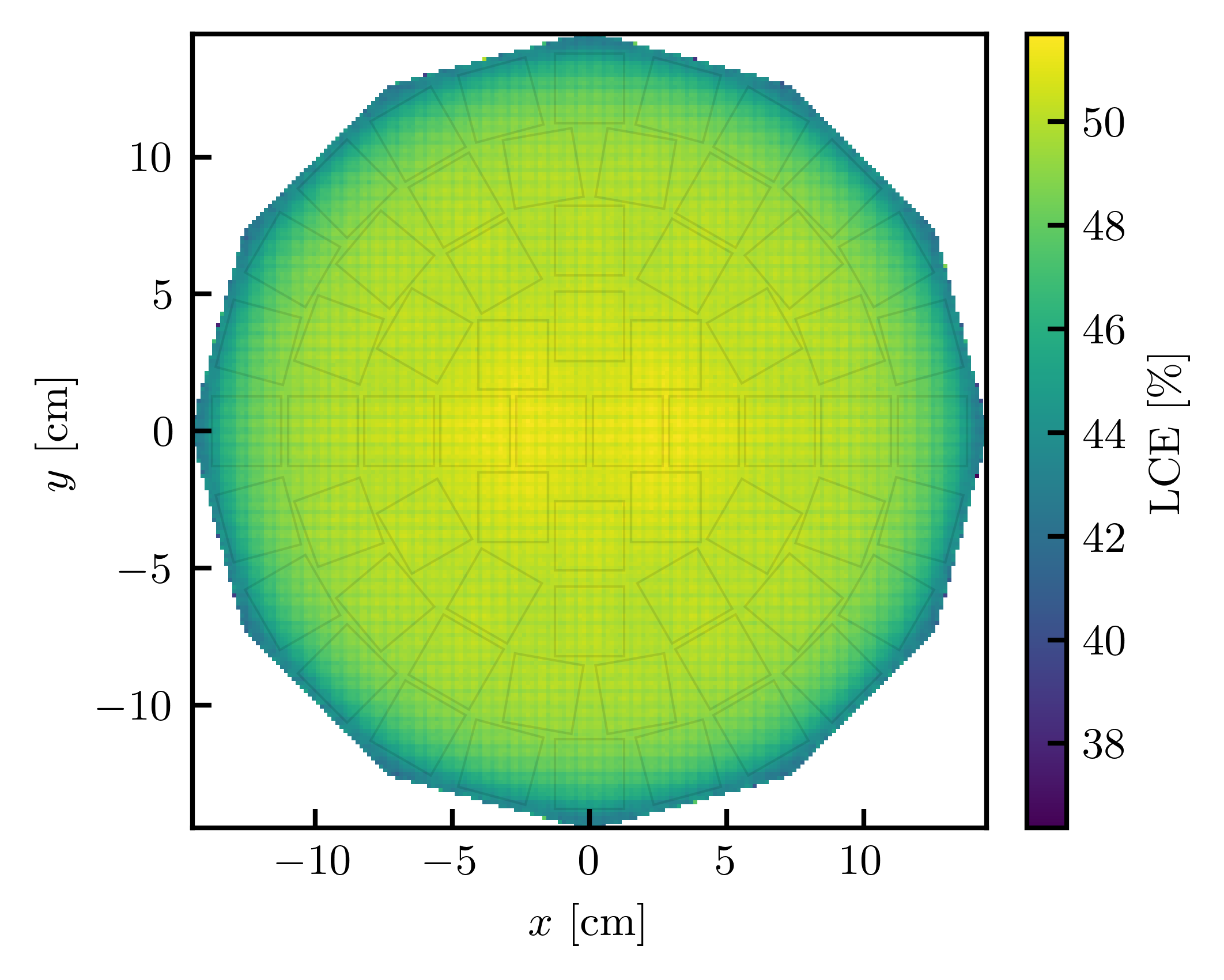}
\caption{The S2 LCE map with both top and bottom PMT arrays. Simulations are performed by generating S2 photons in the GXe region below the anode.}
\label{fig:s2_lce}
\end{figure}

To explore the capability of the RELICS detector on the ionization-only detection channel, the light collection efficiency\,(LCE) of S2 is simulated by \textsl{RelicsSim} with the optical parameters listed in Tab.\,\ref{tab:pars}. The LCE map is shown in figure~\ref{fig:s2_lce}. An averaged $g_2$ value of 30\,PE/e$^{-}$ is estimated based on the S2 LCE map, the properties of the PMTs, such as $\epsilon_{\rm{QE}}$, $\epsilon_{\rm{CE}}$, and $p_{\rm{DPE}}$ that also listed in Tab.\,\ref{tab:pars}, and a 100\% electron extraction efficiency on the liquid surface. An electron lifetime of 1.2\,$\rm m$s is applied for S2 estimation considering the probability of electrons loss to electronegative impurities during their drift. Under these hypotheses, we assume a 100\% trigger efficiency at 4\,e$^-$\,(120\,PE) targeted detection threshold for RELICS sensitivity study.

\begin{table}[htb]
\centering
\caption{\label{tab:pars}The detector related parameters.}
\bigskip
\begin{tabular}{|c|c|}
\hline
\multicolumn{2}{|c|}{\textbf{Optical parameters}} \\
\hline
PTFE reflectivity (LXe \& GXe) &  99\%\,\citep{LUX:2012ybv}      \\
LXe Rayleigh scattering length &  30\,cm\,\citep{Seidel:2001vf}  \\
LXe absorption length          &  50\,m\,\citep{Baldini:2004ph}  \\
Electrodes reflectivity        &  57\%\,\citep{Bricola:2007zz}   \\
\hline
\multicolumn{2}{|c|}{\textbf{Signal generation}} \\
\hline
PMT quantum efficiency ($\epsilon_{\rm{QE}}$)    & 33\%\,\citep{HamamatsuInc} \\
PMT collection efficiency ($\epsilon_{\rm{CE}}$) & 75\%\,\citep{HamamatsuInc} \\
Double PE probability ($p_{\rm{DPE}}$)         & 21.9\%\,\citep{Faham:2015kqa,LopezParedes:2018kzu}  \\
$g_2$ & 30\,PE/e$^{-}$   \\  
\hline
\multicolumn{2}{|c|}{\textbf{Detector operation}}  \\
\hline
Drift field & 500\,V/cm       \\
Electron lifetime & 1.2\,$\rm m$s\,\citep{PandaX-4T:2021bab}   \\
\hline
\end{tabular}
\end{table}

\section{Expected \cevns events}
\label{sec:signal}


The SM of particle physics well predicts the \cevns process --- a neutrino of any flavor scatters off a nucleus via $Z^0$ boson exchange coherently over the whole nucleus at a low momentum transfer ($ Q=\sqrt{2m_t E_R}$). The differential cross-section can be written as Eq.\,\ref{eq_cevns_cs} below:

\begin{equation}
\label{eq_cevns_cs}
\begin{split}
    &\frac{d\sigma}{dE_{R}} = \frac{G^2_F}{2\pi}\frac{Q^2_w}{4}F^2 \left(2m_tE_{R} \right)m_t \left[2-\frac{m_tE_{R}}{E_{\nu}^2} \right]  \\
\end{split}
\end{equation}
where $G_F$ is Fermi's coupling constant, 
$Q_w$ is the weak nuclear charge,
$F$ is the ground state elastic form factor,
$m_t$ is the mass of the target nucleus,
$E_R$ is the nuclear recoil energy and $E_{\nu}$ the incident neutrino energy.
$QR\lesssim1$ is required to ensure coherence, where $R$ is the nuclear radius. This condition can be largely satisfied when $E_{\nu}\lesssim50$\,MeV. 

For neutrinos scattering in a target medium, the expected \cevns event rate can be expressed as Eq.\,\ref{eq_cevns_rate}, where $N_t$ is the number of targets per unit mass, $\phi(E_{\nu})$ is the incident neutrino flux.

\begin{equation}
\label{eq_cevns_rate}
\begin{split}
    &\frac{dN}{dE_{R}} = N_t\int\phi(E_{\nu}) \frac{d\sigma}{dE_{R}}dE_{\nu}   \\
\end{split}
\end{equation}

\begin{figure}[htbp]
\centering 
\includegraphics[width=.45\textwidth]{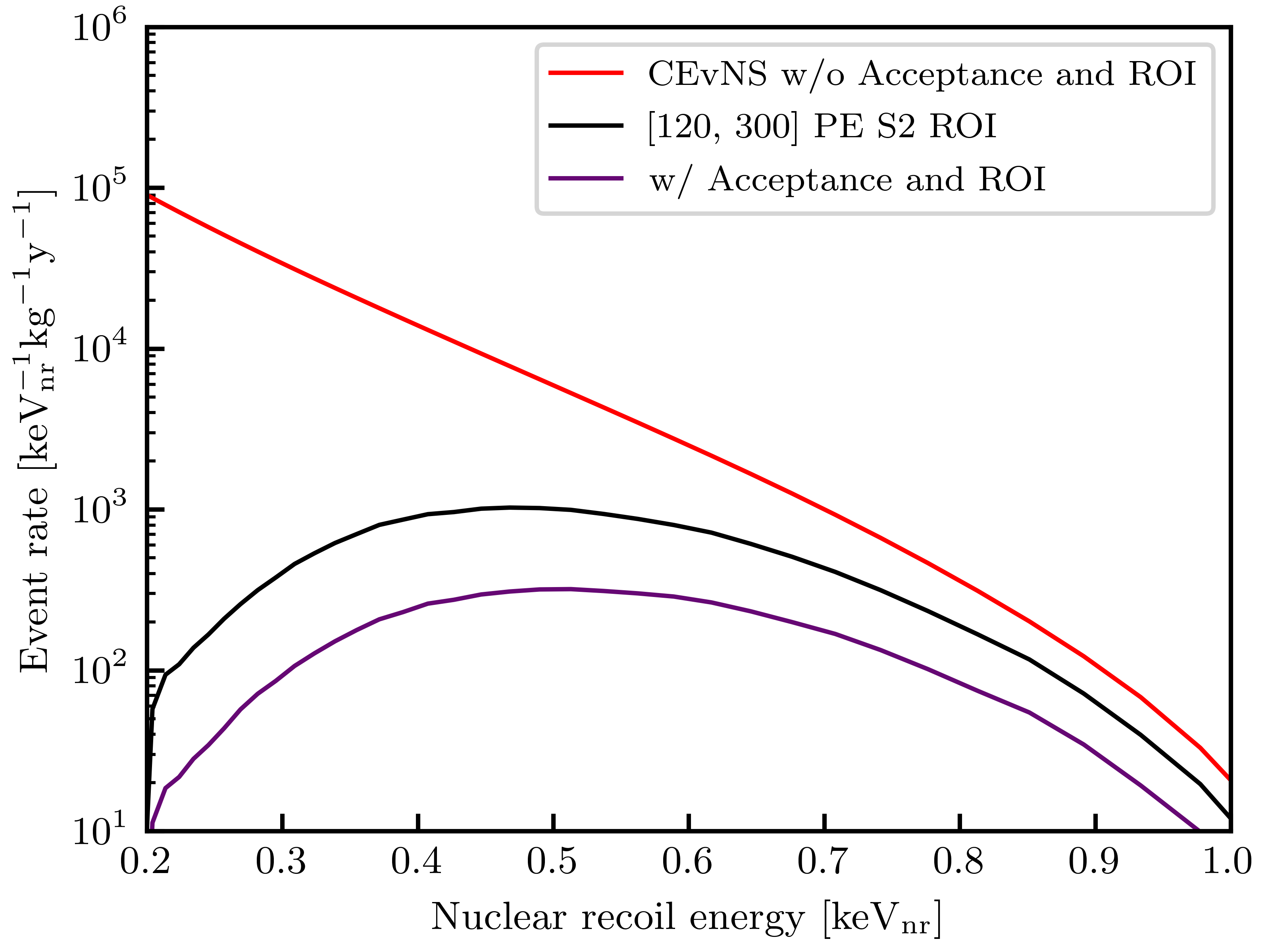}
\caption{The expected \cevns event rate in xenon medium from reactor neutrino with a flux of 10$^{13}$cm$^{-2}$s$^{-1}$. The event rate in RELICS ROI of [120, 300]\,PE with and without signal acceptance are overlaid.}
\label{fig:cevns_event_rate}
\end{figure}

For the reactor neutrinos with energy above 2\,MeV, the Huber-Mueller model\citep{Huber:2011wv, Mueller:2011nm} is adopted for the spectra calculation while P. Vogel’s calculation\,\citep{Vogel:1980bk} is used for below 2\,MeV region, which has never been experimentally explored due to the 1.8\,MeV energy threshold of inverse beta decay reaction. We assume a fuel mixture of $^{235}$U\,(56.1\%), $^{238}$U\,(7.6\%), $^{239}$Pu\,(30.7\%), and $^{241}$Pu\,(5.6\%)~\citep{JUNO:2020ijm}, the total neutrino flux $\phi(E_{\nu})$ can be estimated with a 3\,GW reactor core at 25m distance, and then the \cevns event rate in xenon medium can be obtained accordingly as shown in figure~\ref{fig:cevns_event_rate} (red). The event rate in RELICS region-of-interest\,(ROI) of [120, 300]\,PE is overlaid with (purple) and without (black) signal acceptance, respectively, which will be detailed in Sec.\,\ref{sec:bkgs}. 

\section{Background contributions}
\label{sec:bkgs}

The complexity of the backgrounds is one of the main challenges of reactor \cevns detection. To overcome this challenge, dedicated background simulations have been conducted to optimize the design of the shielding system for RELICS, using \textsl{RelicsSim} and \textsl{RelicsAPT}. The background contributions can be divided into five categories according to their origin, such as (\ref{subsec:cosmic_n_bkgs}) cosmic-ray neutrons, (\ref{subsec:muon_bkgs}) cosmic muons, (\ref{subsec:det_mat}) detector materials, and (\ref{subsec:lxe_bkgs}) internal background from LXe. The last one is (\ref{subsec:instral_bkgs}) delayed electrons, which have been observed by large LXeTPC searching for dark matter and should be treated carefully.  

This section describes all the background components and is summarized in (\ref{subsec:bkgs_summary}). The uncertainties are due to the limited number of simulated exposures. The following criteria are applied to MC data to select the background events that can mimic \cevns signals: 

\begin{enumerate}[label=\roman*.]
\item \textsl{LXe-veto}: The 4$\pi$-LXe veto exhibits a good background suppression because the background events are more likely to deposit energy in both active LXe and 4$\pi$-LXe veto regions. The event will be removed if it contains an NR\,(ER) energy larger than 500\,(100)\,keV in the LXe veto region.

\item \textsl{single-scatter}: Given the small interaction cross-section of neutrino, the probability of a neutrino having more than one interaction in the sensitive LXe volume is negligible. 
We required that the E$\rm_{er}$ must be less than 0.05\,keV$\rm_{er}$ in an NR event, and the second largest energy deposition of E$\rm_{er}$ less than 5\% of the largest one in an ER event.

\item \textsl{FV}: Events are required to be within the center 12\,cm radius fiducial to reduce background at the edge of the TPC, leading to a total fiducial volume (FV) of 32\,kg. 

\item \textsl{S2-width}: For ``S2-only" analysis, the event depth $z$ cannot be accurately estimated. \textsl{S2-width} cut was developed to remove the events around the liquid-gas interface, especially for the backgrounds from detector materials.

\item In addition, dedicated selection methods are developed to suppress backgrounds out of the delayed electrons, such as \textsl{waveform-classifier} and \textsl{pattern-classifier}. These two criteria are described in detail in (\ref{subsec:instral_bkgs}). 

\end{enumerate}

\subsection{Background induced by cosmic-ray neutrons}
\label{subsec:cosmic_n_bkgs}

\begin{figure}[htbp]
\centering 
\includegraphics[width=.45\textwidth]{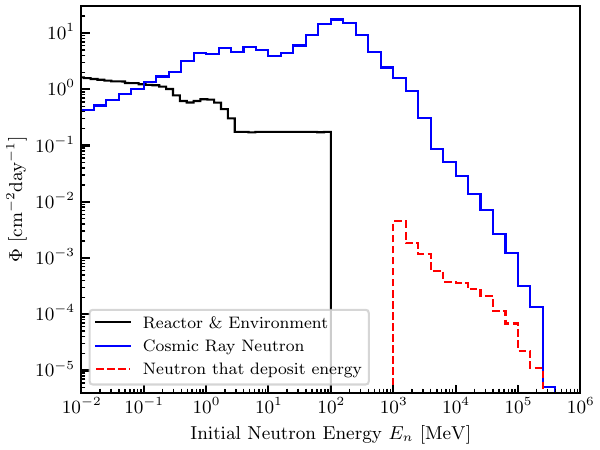}
\caption{The energy spectrum of the CRNs (blue) from CRY calculation. A red dashed line shows the neutrons that can deposit energy in LXe in the \cevns search region. The energy spectrum of the reactor and environmental neutrons from the measurement of CONUS\,\citep{Hakenmuller:2019ecb} experiment is overlaid.}
\label{fig:cosmic_n_spe}
\end{figure}

The cosmic rays that travel through the atmosphere can produce a variety of radiation particles, such as neutrons, protons, pions, etc., through spallation reaction on nitrogen, oxygen, etc\,~\citep{Liu:2021trd, Becchetti:2015nim}. Among them, the high-energy cosmic-ray neutrons (CRNs) can easily pass through the shield and deposit energy in the RELICS detector. The main purpose of the water shield design is to block the CRNs. The size of the water tank is optimized based on the MC simulation. The CRN flux is calculated using the Cosmic-Ray Shower Generator (CRY) source\,~\citep{Hagmann:2007ziw}. Figure\,\ref{fig:cosmic_n_spe} shows the energy spectrum of the CRNs from CRY as the blue curve outside the water shield. Analyzing the MC data shows that only CRNs with energy greater than $\sim$10$^3$\,MeV can deposit energy in LXe, which indicates that the neutrons from the nuclear reactor and laboratory environment with energy less than 10\,MeV\,\citep{Hakenmuller:2019ecb} that also shows in figure\,\ref{fig:cosmic_n_spe}, can not contribute to the backgrounds.

\begin{figure}[htbp]
\centering 
\includegraphics[width=.45\textwidth]{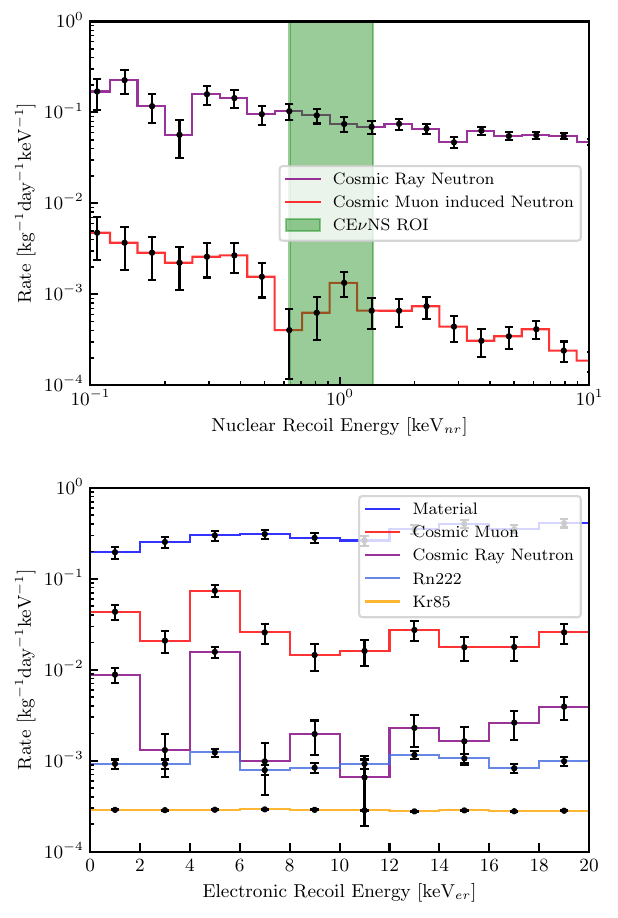}
\caption{Top panel: The NR background distributions from CRNs (dark magenta) and cosmic muons (red). Bottom panel: The ER background distributions from detector materials (blue), CRNs (dark magenta), cosmic muons (red), $^{85}$Kr (sandy brown) and $^{222}$Rn (light blue).}
\label{fig:nr_er_bkgs_kev}
\end{figure}

After processing the MC data, and passing the \textsl{LXe-veto}, \textsl{single-scatter} and \textsl{FV} selections, the CRNs-induced NR background rate is estimated to be (6.0\,$\pm$\,0.6)$\times 10^{-2}\,\rm kg^{-1}day^{-1}$ in $[0.63, 1.36]$\,keV$\rm_{nr}$ ROI of \cevns searching, with the recoil energy spectrum shows in figure\,\ref{fig:nr_er_bkgs_kev} (top). In addition, the secondaries of CRNs will also induce ER background in LXe, with a rate of (3.9\,$\pm$\,0.3)$\times 10^{-3}\,\rm kg^{-1}day^{-1}keV^{-1}$, which is shown in figure\,\ref{fig:nr_er_bkgs_kev} (bottom).

\subsection{Background induced by cosmic muons}
\label{subsec:muon_bkgs}

The RELICS detector will operate without overburden. Hence, special attention has to be paid to the muon-related background. Even though muons can be tagged with 99\% efficiency by the PS veto detector, the neutron background produced by muon interacting with the detector materials is still one of the main background components. Dedicated simulations are performed to evaluate the muon-induced backgrounds. The energy and angular distributions of muons follow the Shukla model\,\citep{Shukla:2016nio}. The energy deposition in LXe that is caused by muon-induced fast neutron and related gammas are recorded by \textsl{RelicsSim}, and then filtered with \textsl{LXe-veto}, \textsl{single-scatter} and \textsl{FV} selections. Assuming a 99\% muon tagging efficiency in 300\,$\mu s$ veto time, the muon-induced NRs and ERs are (0.06\,$\pm$\,0.01)$\times 10^{-2}\,\rm kg^{-1}day^{-1}$ in $[0.63, 1.36]$\,keV$\rm_{nr}$ and (28.0\,$\pm$\,2)$\times 10^{-3}\,\rm kg^{-1}day^{-1}keV^{-1}$ in $[0, 20]$\,keV$\rm_{er}$, respectively. The corresponding spectra are shown in figure~\ref{fig:nr_er_bkgs_kev}.

\subsection{Neutrons and gammas from the detector materials}
\label{subsec:det_mat}

Materials constructing the detector contain radioactive contamination. The selection of the materials requires an extensive radioactivity screening campaign for rare event search experiments. The radioactive contamination for all the materials is considered in the RELICS MC model to evaluate the background contribution from these materials. Currently, the radioactivity of isotopes from XENON100\,\citep{XENON100:2011ufy} is adopted. A dedicated effort is also being carried out to select clean materials for the RELICS experiment, leading to similar radioactivity for most of the materials. 

\begin{figure}[htbp]
\centering 
\includegraphics[width=.45\textwidth]{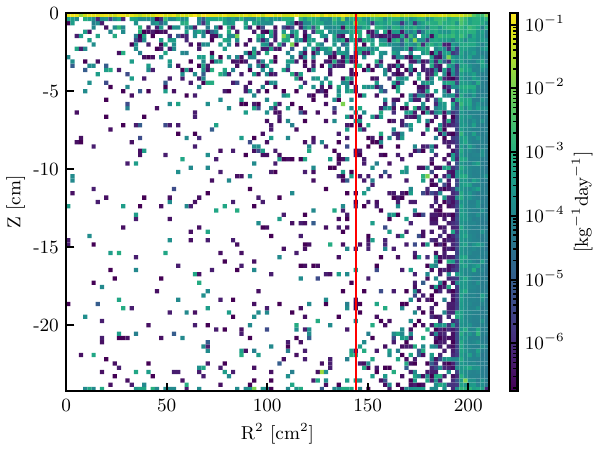}
\caption{Spatial distribution of ER background from detector materials inside the active LXe volume in the energy region of [0, 20]\,keV$\rm_{er}$. The red line indicates the 12\,cm radius fiducial selection.}
\label{fig:er_xyzr}
\end{figure}

The NR background rate induced by neutrons is around one event per year and completely negligible in the \cevns ROI. ER background induced by gammas is studied in this work. The decay of these isotopes is generated and confined uniformly in the corresponding detector component in the simulation. By applying successive selection criteria, including \textsl{LXe-veto}, \textsl{single-scatter} and \textsl{FV} selections, the spatial distribution of ER background events inside the active LXe volume in the energy region of [0, 20]\,keV$\rm_{er}$ is shown in figure\,\ref {fig:er_xyzr}. Figure\,\ref {fig:nr_er_bkgs_kev} (bottom) shows the corresponding energy spectrum, which is more than one order of magnitude higher than other ER background components. The average ER background rate is (310\,$\pm$\,10)$\times 10^{-3}\,\rm kg^{-1}day^{-1}keV^{-1}$ in the FV volume.


\textbf{S2 width selection:} As shown in figure\,\ref {fig:er_xyzr}, a large amount of background events are located close to the liquid-gas interface. These backgrounds are mainly from the radioactivity of PMTs and can not be suppressed with the selection in ``z" in an ``S2-only" analysis. However and as demonstrated in~\citep{XENON:2019gfn}, the pulse shape of S2 signals can discriminate background happening at the liquid-gas interface to signals distributed uniformly in depth ``z". A toy MC was conducted to consider the drift and diffusion properties of the electron clouds in LXe to optimize such a selection. We simulate S2s and their pulse shape parameters with ``z'' distributions shown in figure\,\ref{fig:er_xyzr} with energy in the \cevns ROI. The S2s are simulated via \textsl{RelicsAPT} where the NEST model is incorporated. The S2 pulse shape is simulated using the electron diffusion model $\sigma_e = \sqrt{\frac{2D_Lt}{v_d^2}+\sigma_0^2}$, where $D_L$ is the longitudinal diffusion coefficient, $v_d$ is the electron drift velocity, $\sigma_0$ is the width of single electron waveform\,\citep{Sorensen:2011qs}. In our simulation. $D_L$ = 16\,cm$^2$/s, $v_d$ = 0.174\,cm/$\mu$s, and $\sigma_0$ = 0.19$\mu$s. 

\begin{figure}[htbp]
\centering 
\includegraphics[width=.45\textwidth]{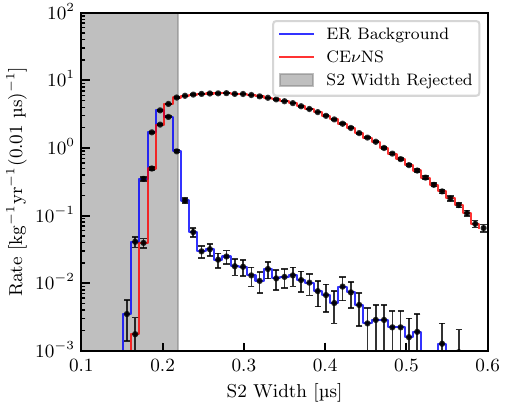}
\caption{The distributions of the S2 width from the uniformly distributed \cevns signal and ER background from detector materials. The dark region is the event that is being rejected by the \textsl{S2-width} criterion.}
\label{fig:s2widthcut}
\end{figure}

Uniformly distributed \cevns signal is simulated using the same procedure. The distributions of the S2 width are shown in figure\,\ref{fig:s2widthcut}, for both \cevns signal and ER background from the detector materials. A threshold of 0.22 $\mu s$ on S2 width can give an average background rejection power of 94\%, with a high \cevns signal acceptance of 86\%.

\subsection{Internal backgrounds in LXe}
\label{subsec:lxe_bkgs}
Radioactivity in LXe can directly deposit energies and produce background in the \cevns ROI. These radioactivities mostly come in the form of noble gases, such as argon, krypton, xenon, and radon. This section describes how these radioactivities are created and evaluates their contribution to the \cevns ROI.  

\subsubsection{Intrinsic $^{222}$Rn and $^{85}$Kr}
\label{subsubsec:rn_kr}

Krypton mixes into xenon gas when it is extracted from the air. $^{85}$Kr, regardless of its low abundance of $2\times10^{-11}$ in natural krypton, is one of the main backgrounds in any LXe experiments searching for rare events. The concentration of krypton in xenon is usually reduced below ppt level in dark matter search experiments, such as XENON1T and PandaX-4T, through distillations with dedicated columns~\citep{XENON:2016bmq,Yan:2021cxp}. In RELICS, we assume a $^{nat}$Kr/Xe concentration of 10\,ppt and expect (0.28\,$\pm$\,0.01)$\times$10$^{-3}\,\rm kg^{-1}day^{-1}keV^{-1}$ as shown in figure~\ref{fig:nr_er_bkgs_kev} (bottom), which is a subdominant component.

The other intrinsic background in LXe comes from the decays of $^{222}$Rn, which emanates from any material touching xenon gas or liquid. The emanated $^{222}$Rn diffuses homogeneously inside the LXe volume within its relatively long half-life of 3.8\,days. The $\beta$ decay of $^{214}$Pb is another important background component in the \cevns ROI in the decay chain of $^{222}$Rn. With similar size of TPC and materials, the radioactivity of  $^{222}$Rn in RELICS is comparable to XENON100. Assuming a conservative $^{222}$Rn radioactivity of 40\,$\rm \mu Bq \cdot kg^{-1}$ as achieved by the XENON100 experiment\,\citep{XENON100:2011ufy} ($\leq$20\,$\rm \mu Bq \cdot kg^{-1}$), the expected ER background rate is estimated as of (0.96\,$\pm$\,0.03)$\times$10$^{-3}$\,kg$^{-1}$day$^{-1}$keV$^{-1}$ as shown in figure~\ref{fig:nr_er_bkgs_kev} (bottom), which is also a subdominant component in the \cevns search.


\subsubsection{Cosmogenic isotopes $^{127}\rm Xe$ and $^{37}\rm Ar$}
\label{subsubsec:cos_iso}

The RELICS experiment, which is to be operated without overburden, is inevitably exposed to cosmic radiation. We consider cosmogenic radioactive isotopes through the activation and spallation of xenon atoms. In particular, $^{127}\rm Xe$ and $^{37}\rm Ar$ are known cosmogenic backgrounds in LXeTPCs\,\citep{PandaX-II:2017hlx, LZ:2022lsv} due to their characteristic x-rays or Auger electrons through electron capture process and their relatively long half-life of 36.4 and 35.0\,days, respectively. 

Natural xenon does not contain $^{127}\rm Xe$. $^{127}\rm Xe$ can be produced from $^{128}\rm Xe$ when cosmogenic muons kick off a neutron in the nucleus with a cross-section of (2.74\,$\pm$\,0.4)$\times10^{-24}$\,cm$^2$\,\citep{Gavrilyuk:2015tva}. Additionally, we consider another production channel through the neutron capture on $^{126}\rm Xe$ with an upper limit cross-section of 0.7$\times 10^{-26} $cm$^2$ at 95\% C.L\,\citep{Gavrilyuk:2015tva}. The production rate of $^{127}\rm Xe$ in LXe is calculated based on its production cross-section, the muon flux\,\citep{Shukla:2016nio} and thermal neutron flux\,\cite{Munehiko:2007}. The $^{127}\rm Xe$ decays to $^{127}\rm I$ via electron capture\,(EC) process, with a half-life of 36.4\,days. We assume that the production and decay of $^{127}\rm Xe$ reach equilibrium before deploying the RELICS detector inside the shielding. The estimated equilibrium decay rate is less than 9.5\,atoms/day/kg. The shielding system of RELICS can attenuate neutrons efficiently, and the equilibrium decay rate can decrease to 0.79\,atoms/day/kg after $\sim$300 days.

Following the EC, the $^{127}\rm I$ is left primarily populating at the excited state of either 375 or 203\,keV, which promptly decays to the ground state with gamma-ray emissions. The EC occurs from either the K, L, M, or N shell electrons and is subsequently filled with electrons from higher levels via emission of x-rays or Auger electrons with total energy (relative intensity) of 32.2\,keV\,(83.4\%), 5.2\,keV\,(13.1\%), 1.1\,keV\,(2.9\%) or 186\,eV\,(0.7\%), respectively. The final background contribution from $^{127}\rm Xe$ is simulated with \textsl{RelicsSim}. The double-scatter criteria can remove most of these background events due to the simultaneous gamma-ray and x-ray or Auger electrons emission. If EC capture is happening from the N-shell electrons with a binding energy of 186\,eV, it may contribute background in the \cevns ROI when the emitted gamma rays, either with energy of 375 and 203\,keV, escape from the LXe detector and the active LXe veto. The event rate from 186\,eV energy deposition is less than $10^{-4}$\,kg$^{-1}$day$^{-1}$ after all the selection criteria described above,  which is negligible for \cevns search.

We consider three possible mechanisms for $^{37}\rm Ar$ production in xenon, following the approaches in\,\citep{LZ:2022kml}. Firstly, the extremely low content of $^{37}\rm Ar$ in the atmosphere, which is generated by spallation of $^{40}\rm Ar$ through $^{40}\rm Ar$(n, 4n)$^{37}\rm Ar$, neutron capture of $^{36}\rm Ar$, or cosmic bombardment of calcium-containing soils, via $^{40}\rm Ca$(n, $\alpha$)$^{37}\rm Ar$. This mechanism only needs to be considered in the context of air leaks of an LXe detector, which is strictly limited to below 0.1 liter per year to ensure the purity level of LXe. Secondly, the spallation of detector materials, especially iron in steel. However, its contribution is greatly suppressed due to the slow diffusion rate of argon in steel\,\citep{LZ:2022kml}. Lastly, the direct spallation of xenon atoms by high-energy cosmogenic protons and neutrons. Only the last possibility is considered in this work. The semi-empirical formula by Silberberg and Tsao was used to calculate the spallation cross-section\,\cite{Silberberg:1973jxa}. The CRY is used to calculate the proton spectrum. Gordon’s neutron spectrum\,\citep{Gordon:2004non} is adopted for cosmogenic neutron estimation. The production rate of this process is estimated to be 0.024\,atoms/kg/day with an uncertainty of $\sim$100\%\,\citep{LZ:2022kml}. The produced $^{37}\rm Ar$ decays to the ground state of $^{37}\rm Cl$ via EC and subsequently deposits energy via x-rays or Auger electrons. The total energy (relative intensity) for K, L, and M shells are 2.82\,keV\,(90.2\%), 0.270\,keV\,(8.9\%), and 0.018\,keV\,(0.9\%), respectively. As a results, the background rate of $^{37}\rm Ar$ in the \cevns ROI is below $10^{-4}$\,kg$^{-1}$day$^{-1}$ and completely negligible for \cevns search.

\subsection{Delayed electrons}
\label{subsec:instral_bkgs}

Besides the background components induced by real physical sources, another type of background must be considered in the \cevns ROI. This background is caused by a high emission rate of delayed electrons after S2 signals in the TPC, which could be incorrectly identified as \cevns candidate in ``S2-only" analysis. Such background has been observed by large LXeTPCs experiments searching for dark matter particles in deep underground laboratories, such as XENON1T\,\cite{XENON:2021qze}, PandaX-4T\,\cite{PandaX:2022xqx} and LUX\,\cite{LUX:2020vbj}. The source of this background is not well understood yet, but we can roughly divide the delayed electrons into two groups.

The first kind of delayed electrons is produced via photo-ionization of impurities or metals in the drift region of the TPC. Photo-ionized delay electrons only appear shortly (within a time range that is comparable to the largest drift time of the LXeTPC) after the high energy S2 signals and will be completely subdominant when a 1\,$ms$ exposure is rejected for each of the high energy S2s, leading to an exposure loss of $\simeq$1\%.

Besides photo-ionization, delayed electrons can also be produced in a larger time range of up to a few seconds. The rate of delayed electrons is observed to be correlated with the level of impurities in LXe and the amplitude of the extraction field. Muons passing through the LXeTPC result in large energy deposition, generating a substantial number of ionization electrons. These delayed electrons may occur after the primary S2 signal of muon events, leading to a prolonged tail of signals composed of a single or a few electrons. Due to the high rate of cosmic muons, RELICS faces a major challenge in suppressing background from accidental pile-ups of delayed electrons in a short time range that is comparable to the width of a physical S2 signal. This background is hereafter referred to as Delayed Electron (DE).

\subsubsection{Expected rate of DE}
\label{subsubsec:delayed_electron}

As shown in previous studies with XENON1T\,\cite{XENON:2021qze} and LUX\,\cite{LUX:2020vbj}, the emission rate of delayed electrons are correlated, in time and position, with the previous high-energy events. This part of the background is referred to as a correlated delayed electron background and can be significantly suppressed due to its unique correlation with proceeding high-energy events. We simulate muons passing through the TPC active volume using Geant4, tracking their interaction vertex and the energy deposited along the track. The ionization signals are generated along the muon track, assuming a constant ionization yield of 50 electrons per keV. We then simulated the generation of delayed electrons, assuming that 0.1\% of the total ionization electrons from primary interactions are delayed emissions in a time range of 0.1\,ms to 2\,s, comparable to \cite{XENON:2021qze}. The distribution in ``delayed time" (time with respect to the primary S2) for delayed electrons is assumed to follow a power law with a coefficient of -1.1, as shown in XENON1T\,\cite{XENON:2021qze}. These delayed electrons' horizontal positions\,($x,\,y$) are assumed to be the same as their primary S2s. Position reconstruction resolution for single electron S2s, which is expected to be 20\,mm from optical simulations, is taken into account in this analysis.
Ionization S2 signals are grouped together if the time gap between them is below 1\,$\mu$s in order not to split S2 waveforms of events happening at the bottom part of the TPC. As a result, the pile-up of DE (called Pile-up DEs) leads to a significant background in \cevns search. In RELICS, we expect more than 10\,Hz of muon events in the LXeTPC, leaving a large amount of single electron S2 signals. The expected event rate of a single-electron pile-up background is summarized in Tab.\,\ref{table:pileuprate}. 
Without implementing appropriate cuts to reduce the rate of these pile-up events, \cevns signals with 3 to 6 electrons will be completely overwhelmed. 

\begin{table}[htb]
\centering
\caption{A comparison of the \cevns signal and the Pile-up DEs background in one year of exposure with 32\,kg of fiducial mass.}
\bigskip
\begin{tabular}{|c|c|c|}
\hline
\quad N$_e$ \quad &  \qquad \cevns \qquad\qquad   &  \qquad Pile-up DEs \qquad\qquad \\ 
\hline
 3 &  2.4$\times 10^{4}$  &  9.8$\times 10^{8}$ \\
\hline
 4 &  1.1$\times 10^{4}$  &  7.3$\times 10^{6}$ \\
\hline
 5 &  5.2$\times 10^{3}$  &  6.9$\times 10^{5}$ \\
\hline
 6 &  2.4$\times 10^{3}$  &  7.5$\times 10^{4}$ \\
\hline
\end{tabular}
\label{table:pileuprate}
\end{table}

\subsubsection{Waveform classifier}
\label{subsubsec:wf_discri}

\begin{figure}[htbp]
\centering 
\includegraphics[width=.45\textwidth]{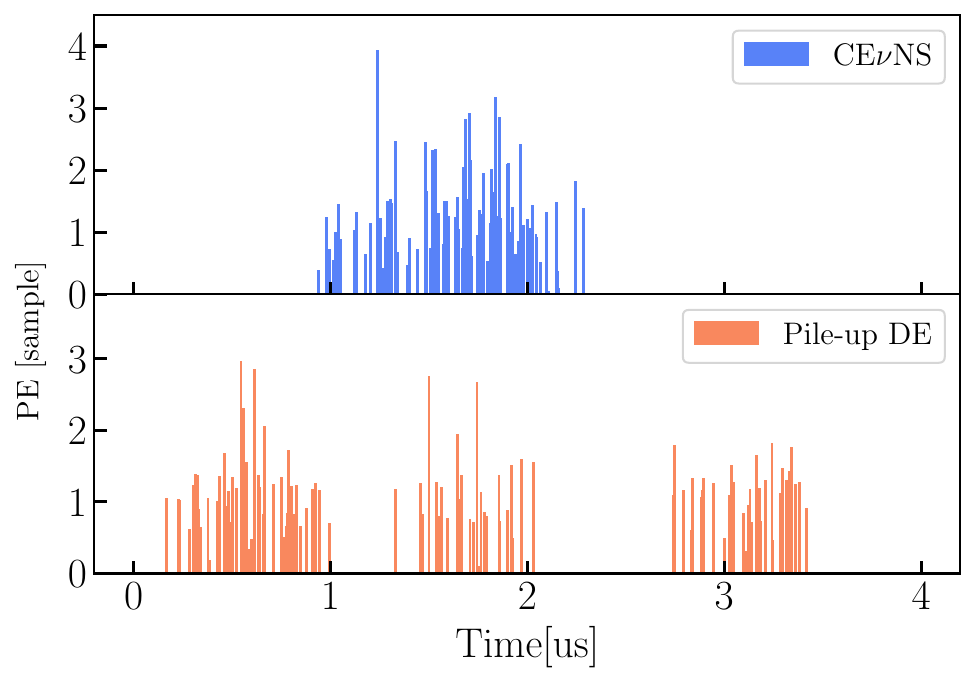}
\caption{Simulated waveform of Pile-up DE and CE$\nu$NS. Recording the number of photo-electrons collected for each sample, the upper panel corresponds to real physical events, and the lower panel represents the Pile-up DE events.}
\label{fig:waveform}
\end{figure}

\begin{figure}[htbp]
\centering 
\includegraphics[width=.45\textwidth]{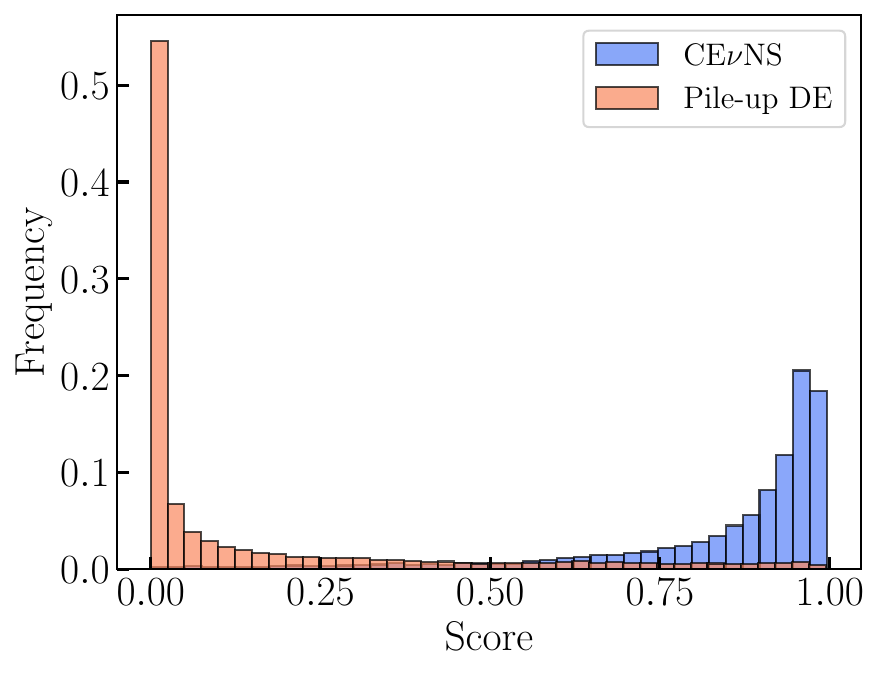}
\caption{Result of waveform discrimination by CNN. The waveform is evaluated by CNN, which provides predicted scores indicating the discriminability between the two types of waveform.}
\label{fig:scores}
\end{figure}

The waveform characteristics of Pile-up DE events differ from the physical interactions in the LXeTPC, such as \cevns and other backgrounds. In physical interactions, electrons drift upwards under the drifting electric field, resulting in signal waveform distributions that follow a Gaussian distribution in time. As shown in figure\,\ref{fig:waveform}, Pile-up DEs are distinguishable from physical interactions in terms of the shape of waveforms. 
The discrimination between them is maximized by employing a convolutional neural network (CNN), as shown in figure \,\ref{fig:scores}. While maintaining $\sim$80\% of the physical interactions with a score over 0.8, the Pile-up DE events are reduced by approximately one order of magnitude.

\subsubsection{Pattern classifier}
\label{subsubsec:lp_spacetime}

Beyond the distinctions in time distribution of the waveform, the planar distribution of signals within the top PMT arrays also presents notable differences between Pile-up DE and physical interactions. \cevns and other physical signals are composed of multiple electrons generated from the same position below the liquid surface in the detector, while the DE signals are mostly generated around the previous muons' tracks. These DEs are merged together into larger signals only due to their temporal proximity. Therefore, the distributions of photo-electrons from \cevns signals and DE signals differ in the top PMT array of the detector. Real physical interactions appear more like a superposition of multiple single-electron signals from the same location, whereas the Pile-up DEs are formed by the superposition of single-electron signals originating from multiple different horizontal positions, thus appearing more dispersed.

For an \cevns or physical interaction, the expected photo-electrons detected in each channel on the top PMT array are denoted as $\lambda_{ch}$. $\lambda_{ch}$ is determined by the total area of S2, the relative LCE in that channel, and the PMT responses including detection efficiency. LCE of each PMT is derived with light propagation simulation in the RELICS TPC. It is assumed that the number of photo-electrons received by the PMT channel follows a Poisson distribution with $\lambda_{ch}$ as the mean. The detected photo-electrons are further derived considering the resolution of PMTs to a single photo-electron.

While the ``expected" $\lambda_{ch}$ is simulated with multiple electrons generated from the same position, the ``observed" photo-electrons $N^{pe}_{ch}$ in each channel from an ``unknown" event is calculated by summing up the simulation from each electron at its corresponding position. Given the expected and observed distribution pattern, we calculated the goodness-of-fit likelihood to determine whether the event is from physical interactions or Pile-up DEs. The logarithm of goodness-of-fit likelihood $P$ is written as:

\begin{equation}
\label{LogLikelihoodCalculation1}
    P = \sum_{ch=0}^{63} \log \left(\frac{{\lambda_{ch}}^{N^{pe}_{ch}}\times e^{-\lambda_{ch}}}{{N^{pe}_{ch} !}} \right)
\end{equation}

\begin{figure}[htbp]
\centering
\includegraphics[width=.4\textwidth]{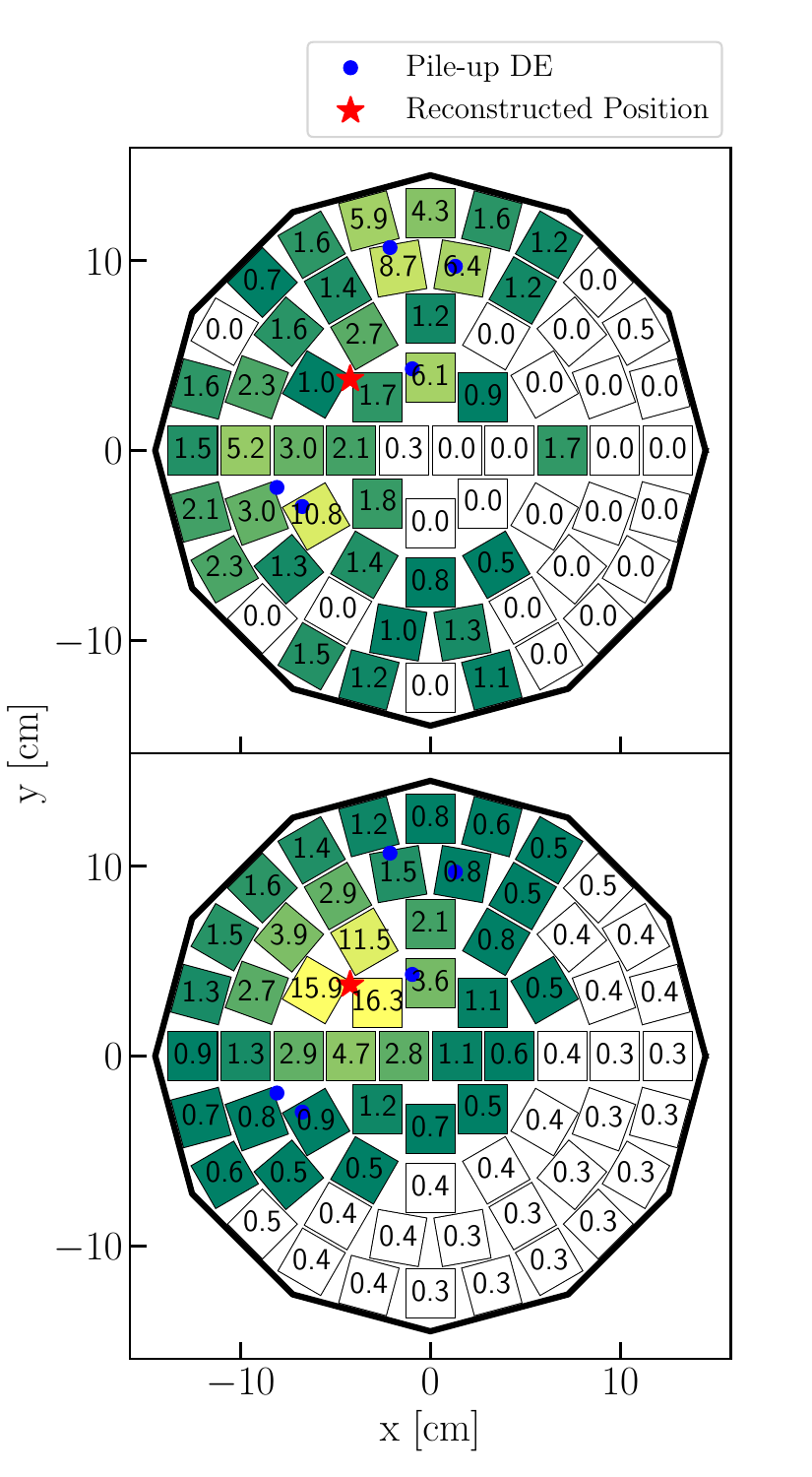}
\caption{The observed S2 pattern ($N^{pe}_{ch}$) on top PMT array from one Pile-up DE event (top) and the expected pattern ($\lambda_{ch}$) from the corresponding reconstructed position (bottom) with the same S2 area from physical interaction. Expected PE numbers for each channel are provided. Only channels with an expected area greater than 0.5\,PE are shown with color.}
\label{fig:S2_pattern}
\end{figure}

The S2 distribution pattern on top PMT array of the Pile-up DEs rejected by a selection in $P$ (figure~\ref{fig:S2_pattern},\,top) would look like that of physical interactions (figure~\ref{fig:S2_pattern},\,bottom). Since both the horizontal position and emission time of DEs are strongly correlated to the previous muon tracks, we further calculate the ``Space-time Correlation Coefficient" of each event to prior muons to distinguish Pileup DEs from physical interactions that are independent of any muon events in the RELICS LXeTPC. 

For each energy deposition of muon passing through the RELICS TPC, the reconstructed position of DEs would follow a Gaussian distribution determined by the resolution of position reconstruction. The distribution of DEs in time with respect to the previous muon events is assumed to follow a power law, as seen in \cite{XENON:2021qze}. Assuming $N_i$ is the expected number of DE in a specific position, the probability distribution function of DEs can be expressed as:

\begin{equation}
\label{SpaceCorrelation}
    P_{i} =  N_i \times \left(t-t_i\right)^{-\gamma}\\
     \times \frac{1}{2 \pi \cdot \sigma^2} e^ {-\frac{\left(x-x_{\mathrm{i}}\right)^2+\left(y-y_{\mathrm{i}}\right)^2}{2\sigma^2}}
\end{equation}
where $(x_i,y_i)$ is the position of muon track,  $(x,y)$ is the observed position, $t-t_i$ is the time delay to the muon track and $\sigma$ is the resolution of reconstructed position. Considering that the emission of DE can last up to $\simeq$2s and the muon event rate of $\simeq$10\,Hz, we sum up the contributions of $P_i$ from the previous 20 muons and define the log of this value as the ``Space-time Coefficient", which is defined as:

\begin{equation}
\label{SpaceTimeCorrelation}
    C_{Space-time} = \log \left( \sum_{i} P_i\right)
\end{equation}

\begin{figure}[htbp]
\centering 
\includegraphics[width=.45\textwidth]{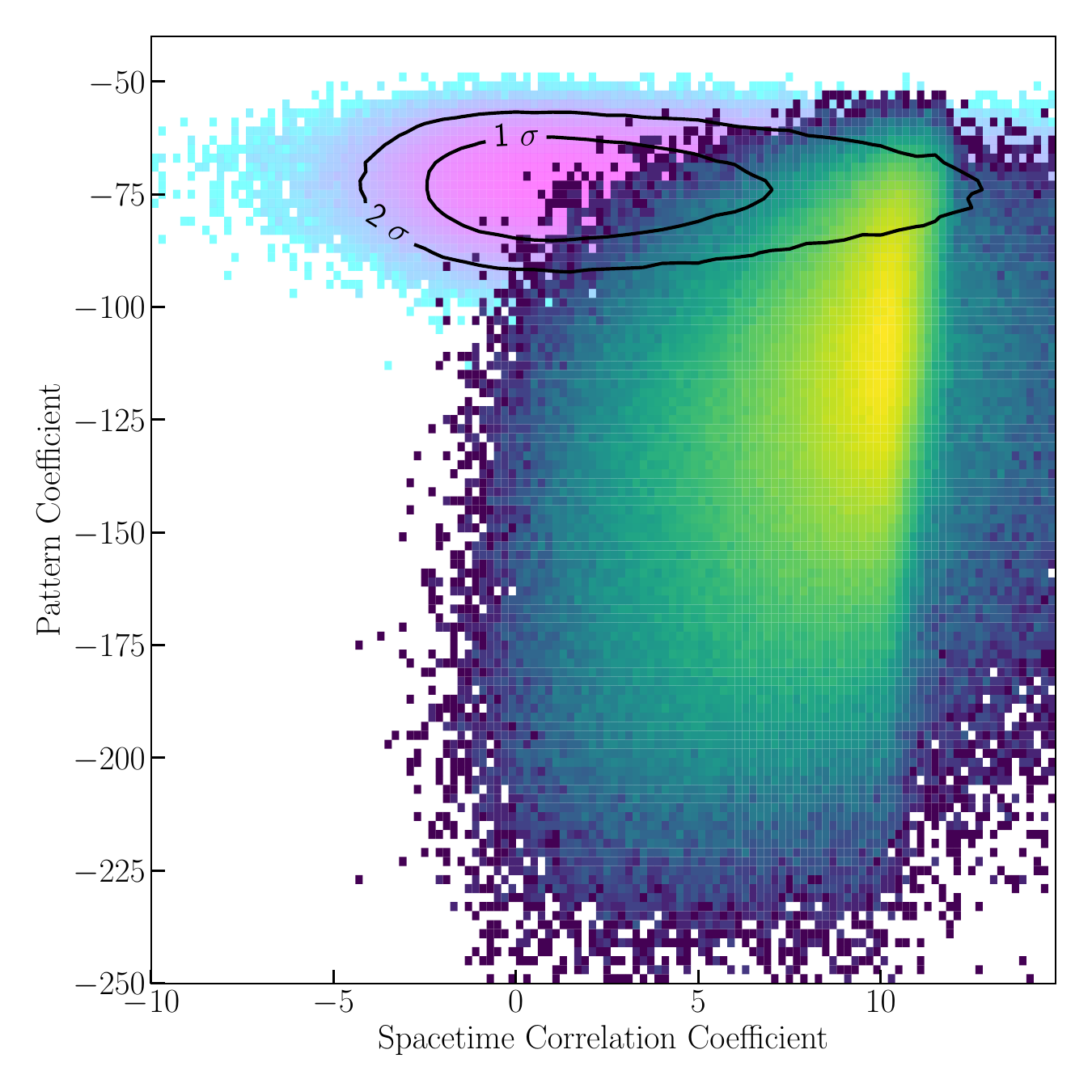}
\caption{Distribution of pattern likelihood and space-time correlation of Pile-up DE and \cevns S2 signals with S2 area range from 150 to 160\,PE. \cevns events are represented by green/yellow points and Pile-up DE events are represented by blue/violet points. Contours of \cevns events distribution are also overlaid.}
\label{fig:2d_dis}
\end{figure}

We simulated the Pile-up DE events of the detector operating for 8500 hours and calculated the ``Pattern coefficient" $P$ and ``Spacetime Correlation Coefficient" $C_{Space-time}$ for each event. 
The event distribution in ($P$, $C_{Space-time}$) space slightly varies with different S2 sizes. To clearly illustrate the discrimination power, the Pile-up DE and CE$\nu$NS events in the energy slice of [150, 160]\,PEs in the ($P$, $C_{Space-time}$) space are shown in figure~\ref{fig:2d_dis}. 


Considering the correlation between $C_{Space-time}$, P and the S2 area of each event, we provide a score for each event by Eq.\,\ref{PatternScore}. Events with higher scores are inclined to be from physical interactions.

\begin{equation}
\label{PatternScore}
    Score =  P - \left( k_{st} \cdot C_{space-time} \right) + k_{area} \cdot S2_{area} + b
\end{equation}

While restricting the acceptance of Pile-up DE events lower than $0.005$\%, we scanned the parameter space of $k_{st}$ and  $k_{area}$, in order to retain as high signal acceptance as possible. The fitted values are $k_{st}$ of 5.6 , $k_{area}$ of 0.8 and $b$ of -37.42. 

\begin{figure}[htbp]
\centering 
\includegraphics[width=.45\textwidth]{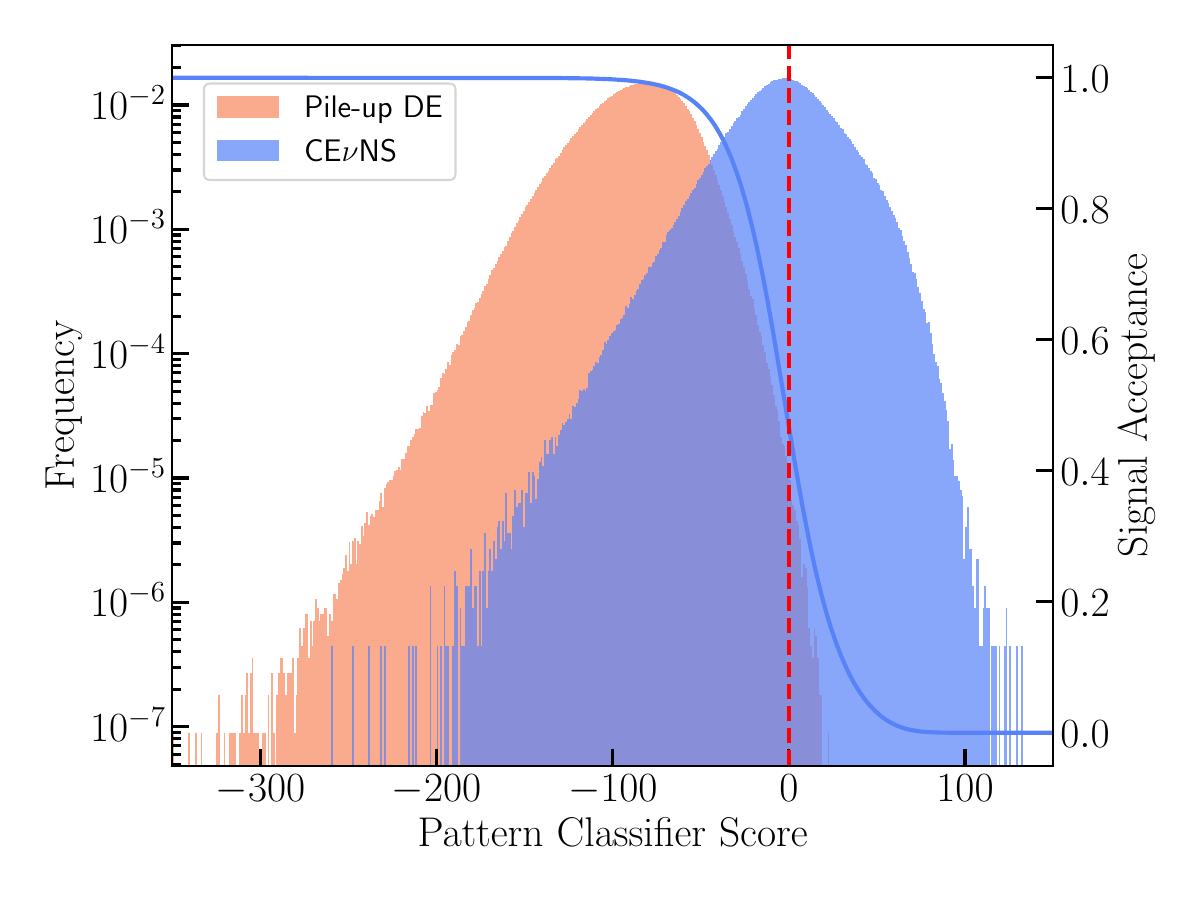}
\caption{Distribution of scores from pattern classifier. The survival rate for \cevns within ROI with scores over 0 is 46.5\% and 0.005\% for Pile-up DE events within ROI.}
\label{fig:pattern_score}
\end{figure}

Figure~\ref{fig:pattern_score} delineates the score distribution for Pile-up DEs and physical interactions. The final classifier identifies all events with scores exceeding 0 as real physical interactions. In the ROI of S2 ranging from 120 to 300\,PE, the acceptance for physical interactions such as \cevns is 46.5\%. This cut reduced the rate of Pile-up DEs by 4 orders of magnitude in the same ROI, leaving the rate of Pile-up DE background below the expected \cevns signal in the whole ROI.

\subsection{Summary}
\label{subsec:bkgs_summary}

\begin{figure}[htbp]
\centering 
\includegraphics[width=.48\textwidth]{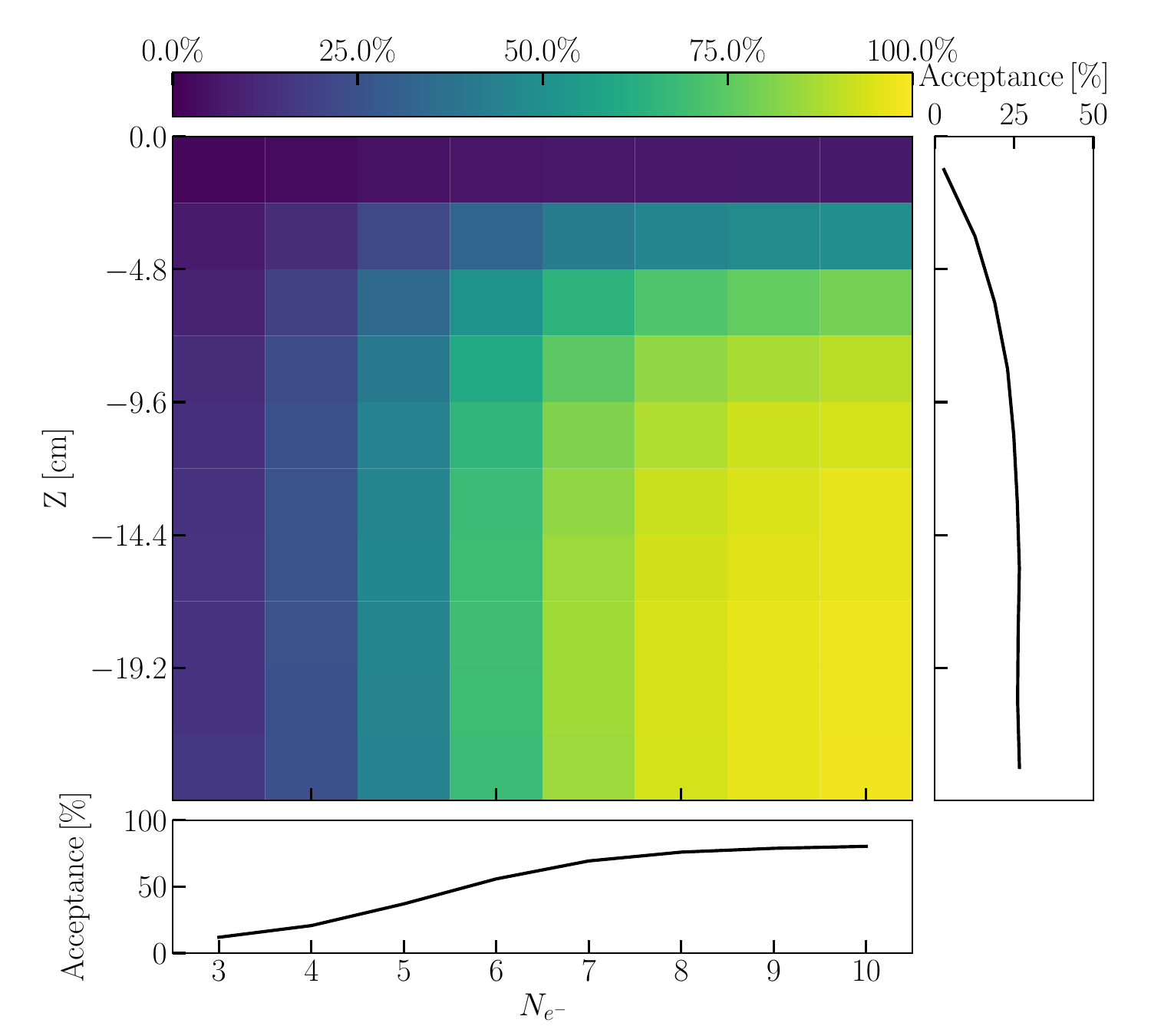}
\caption{The acceptance of \cevns signals by applying the event selection criteria of
\textsl{S2-width}, \textsl{waveform-classifier} and \textsl{pattern-classifier}, in-depth ``z" \textsl{vs} number of electrons $N_{e^-}$ space.}
\label{fig:cevns_accp}
\end{figure}

Considering the correlations between \textsl{S2-width}, \textsl{waveform-classifier} and \textsl{pattern-classifier} criteria, the total acceptance of these three cuts is estimated as shown in figure~\ref{fig:cevns_accp}. 

\begin{figure}[htbp]
\centering 
\includegraphics[width=.45\textwidth]{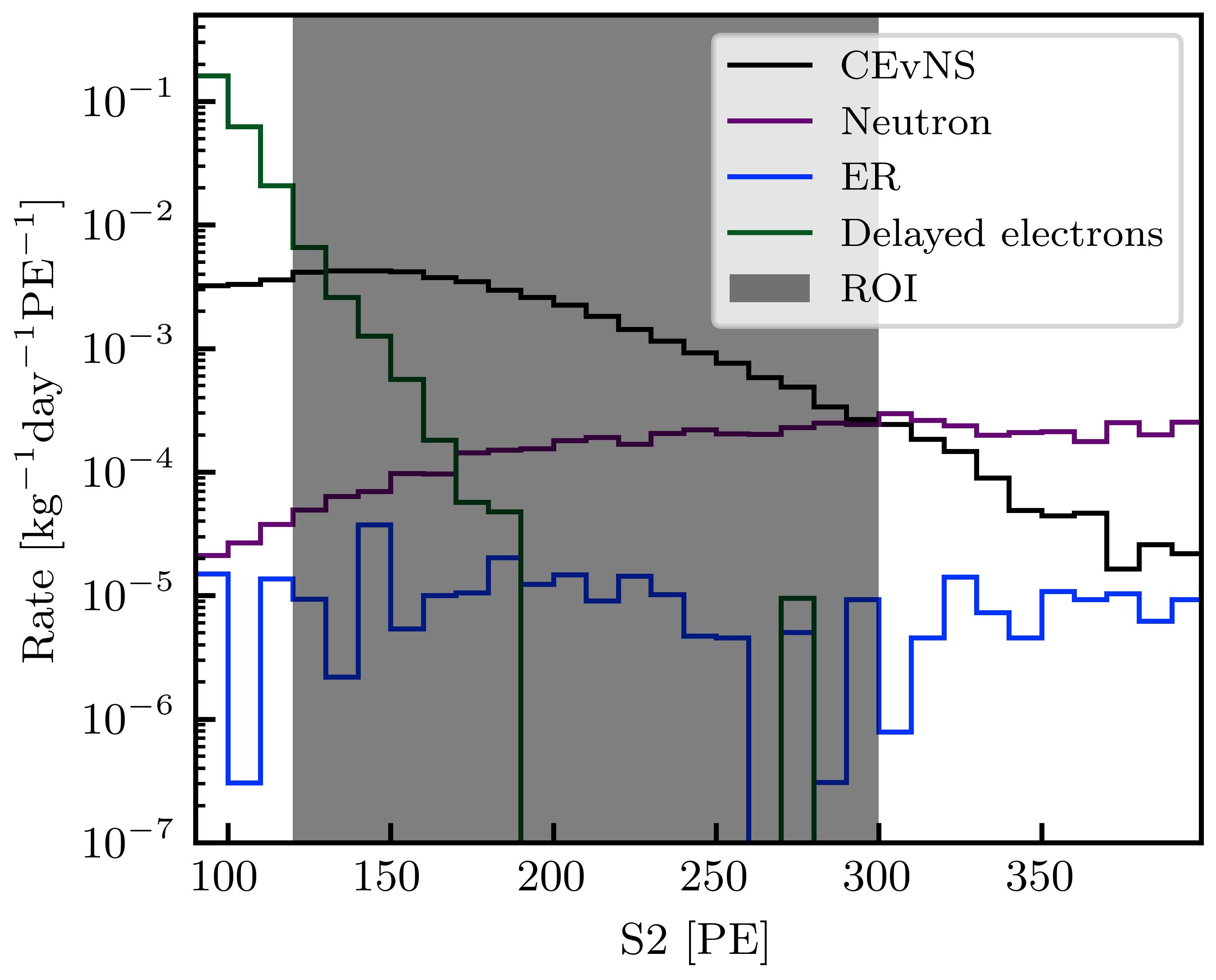}
\caption{Expected S2 distribution of \cevns (Black) signal among all background contributions in the RELICS experiment. The dominant background is the Pile-up DE (Green) induced by muons in the TPC. The second largest background is neutrons (Purple), mainly from the CRNs. The ER from radioactivity of detector material, CRNs, and muons (Blue) is a non-dominant background. The ROI for the \cevns search is defined as 120\,PE to 300\,PE, denoted as the shaded region.}
\label{fig:total_bkgs}
\end{figure}

The distribution of all the background components is converted to S2 size in photo-electrons from recoil energy using \textsl{RelicsAPT}, which is described in Sec.\,\ref{sec:sim_framework}, and summarized in figure~\ref{fig:total_bkgs}. 
The \cevns event rate is overlaid taking into account the signal acceptance. The dominant background contribution comes from the Pile-up DEs, despite its large discrimination power to physical events as described in Sec.\,\ref{subsec:instral_bkgs}. The NR background contribution from CRNs is the second major contribution. The detector materials contribute a relatively low and flat component in ER. The background components' event rate in 32\,kg$\cdot$yr exposure is summarized in Tab.\,\ref{table:bkgs_sum}. 

While not utilized for our baseline sensitivity, we are aware that the reactor-OFF period can be used to further constrain our background model~\citep{KamLAND:2013rgu,Haghighat:2018mve}. Since the reactor-related background is negligible in the total budget, this data will provide a background-only measurement and dedicated runs can be planned in the future to enhance the signal sensitivity.

\begin{table}[htb]
\centering
\caption{The total number of expected events (N$\rm_{evts}$) of all the background components in [120, 300]\,PE ROI of \cevns detection in 32\,kg$\cdot$yr exposure.}
\bigskip
\begin{tabular}{|c|c|}
\hline
Event Type       &  \qquad N$\rm_{evts}$ \qquad\qquad  \\ 
\hline
\hline
 Pile-up DE  &    1325.1         \\
\hline
 CRNs        &        339.9      \\
\hline
Muon-induced neutrons &  1.7  \\
\hline
ER                   &    21.1 \\
\hline
Total Background       &   1687.8  \\
\hline
\hline
\cevns       &       4639.7      \\
\hline
\end{tabular}
\label{table:bkgs_sum}
\end{table}

\section{Physics potential}
\label{sec:physics}

To explore the physics potential of the RELICS detector, this section will discuss its sensitivity to the weak mixing angle and non-standard interactions (NSI) with neutrino has vector couplings to the up and down quarks based on the background estimation in section\,\ref{sec:bkgs} with 32\,kg$\cdot$yr exposure. Several systematic uncertainties are considered, including 5\% reactor neutrino flux, 10\% ER and 5\% NR background uncertainty. The uncertainty of the Q$_y$ introduced in Sec.\,\ref{subsec:lxe_resp} is the dominant one in our estimation.

\begin{table}
\centering
\caption{\label{tab:thets_s2_uncert}List of nuisance parameters shared in the likelihoods for measuring the weak mixing angle and searching for NSI. All the nuisance parameters are constrained by Gaussian distribution, and their standard deviations (constraints) are also listed in the table.}
\bigskip
\begin{tabular}{|c|c|c|}
\hline
\multicolumn{2}{|c|}{Nuisance Parameters} & Constraints\\
\hline
Charge yield morpher & $t$ & 1\,\citep{XENON:2020gfr} \\
Reactor neutrino flux multiplier & $\gamma_f$ & $\sigma_{\gamma_f}=0.05$ \\
ER background rate multiplier & $\beta_1$ & $\sigma_{\beta_1}=0.10$  \\
Neutron background rate multiplier & $\beta_2$ & $\sigma_{\beta_2}=0.05$  \\
Pile-up DE background rate multiplier & $\beta_3$ & $\sigma_{\beta_3}=0.10$  \\
\hline
\end{tabular}
\end{table}

\subsection{Weak mixing angle}
\label{subsec:cs_wma}
A precise measurement of \cevns cross-section allows for a constraint on the weak mixing angle $\theta_\mathrm{w}$. Also known as the Weinberg angle, $\theta_\mathrm{w}$ is conventionally used in the form of $\sin^2{\theta_\mathrm{w}}$ in the comparison between theory and experiment. Since $Q_w$ is given by $N-\left(1-4\,\rm sin^2\theta_\mathrm{w} \right)Z$, $\sin^2{\theta_\mathrm{w}}$ can be extracted from the measured cross-section, and any deviation from the SM prediction will indicate new physics.
  
\begin{equation}
\begin{aligned}
\label{eq:thets_s2_ll}
\footnotesize
\begin{split}
\chi^2 = & \sum_j\left[\frac{\left(N_{\mathrm{obs},j} - N_{\mathrm{exp},j}(\sin^2\theta_W, t)(1 + \gamma_f) 
    - \sum_{i}^3 B_{i,j}(1 + \beta_i)\right)^2}{\sigma^2_{\mathrm{stat},j}}\right] \\
& + t^2 + \sum_{i}^3(\frac{\beta_i}{\sigma^2_{\beta_i}})^2 + (\frac{\gamma_f}{\sigma^2_{\gamma_f}})^2
\end{split}
\end{aligned}
\end{equation}

The $\chi^2$ of likelihood for $\sin^2{\theta_w}$ measurement is defined as Eq.\,\ref{eq:thets_s2_ll} where index $j$ indicates different S2 intervals. $N_{\mathrm{obs},j}$ and $N_{\mathrm{exp},j}$ are the observed and expected signal events in each S2 interval, respectively. $B_{i,j}$ is the background rate estimation and index $i$ runs over background components, including ER and neutrons. 
The rest are nuisance parameters. $\beta_1$, $\beta_2$ and $\beta_3$ are the multipliers of the ER, neutron and Pile-up DE backgrounds, respectively. $t$ denotes the morphers of charge yields. A Gaussian constraint term is imposed on each nuisance parameter, as summarized in Tab.~\ref{tab:thets_s2_uncert}.

\begin{figure}[htbp]
\centering 
\includegraphics[width=.45\textwidth]{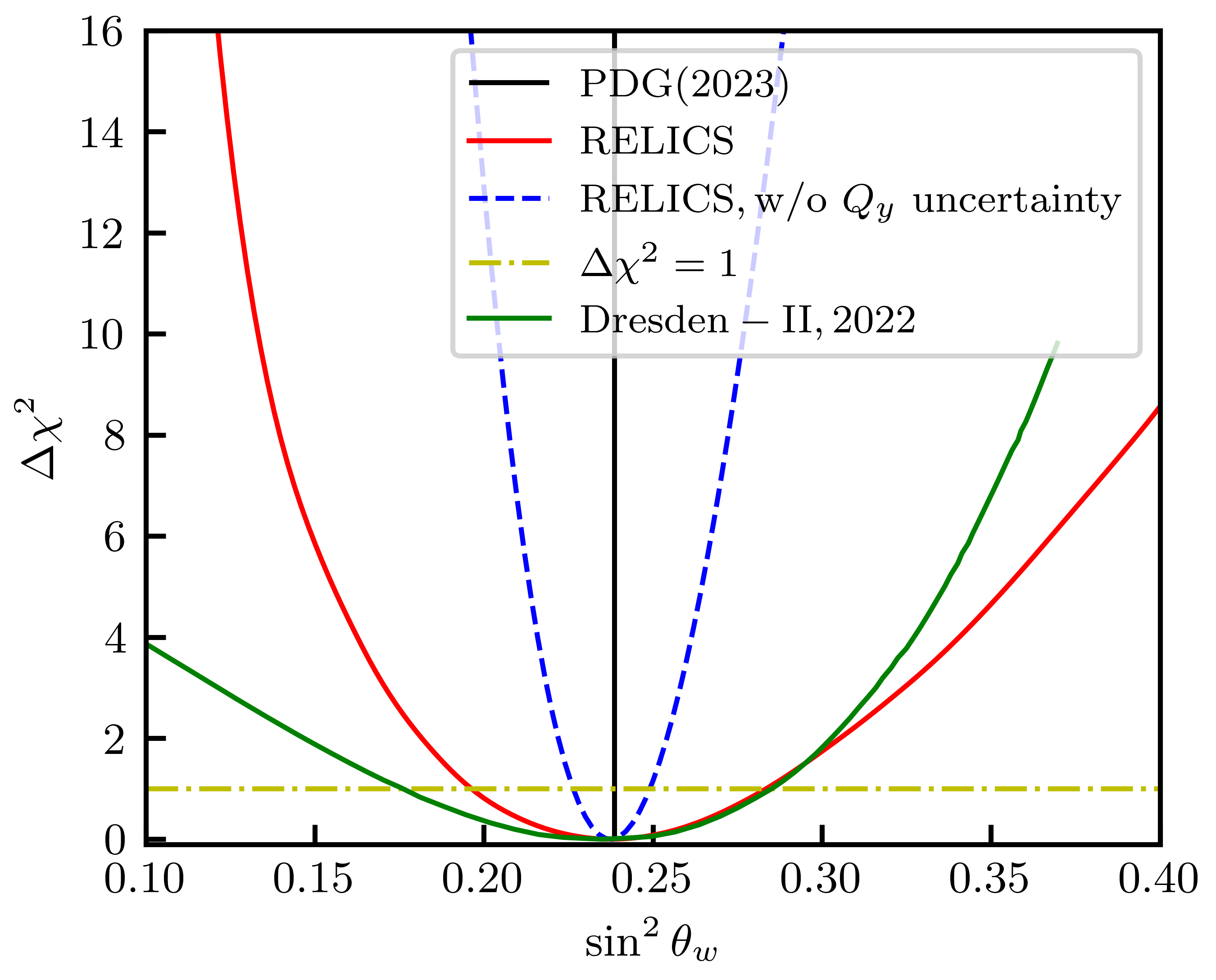}
\caption{The expected constraints on $\sin^2{\theta_w}$ with RELICS. The red line shows the nominal likelihood curve defined by $\chi^2$ in Eq.~\ref{eq:thets_s2_ll}, while the blue dashed line illustrates the likelihood curve with the charge yield fixed. The faint green dot-dashed line marks $\Delta\chi^2=1$, and its intersections with each likelihood curve represent the constraint. The Dresden-II result~(green)\,\citep{Majumdar:2022nby} and the PDG value~(black)\,\cite{ParticleDataGroup:2022pth} are overlaid for reference.}
\label{fig:nu_weak}
\end{figure}

\begin{figure}[htbp]
\centering 
\includegraphics[width=.45\textwidth]{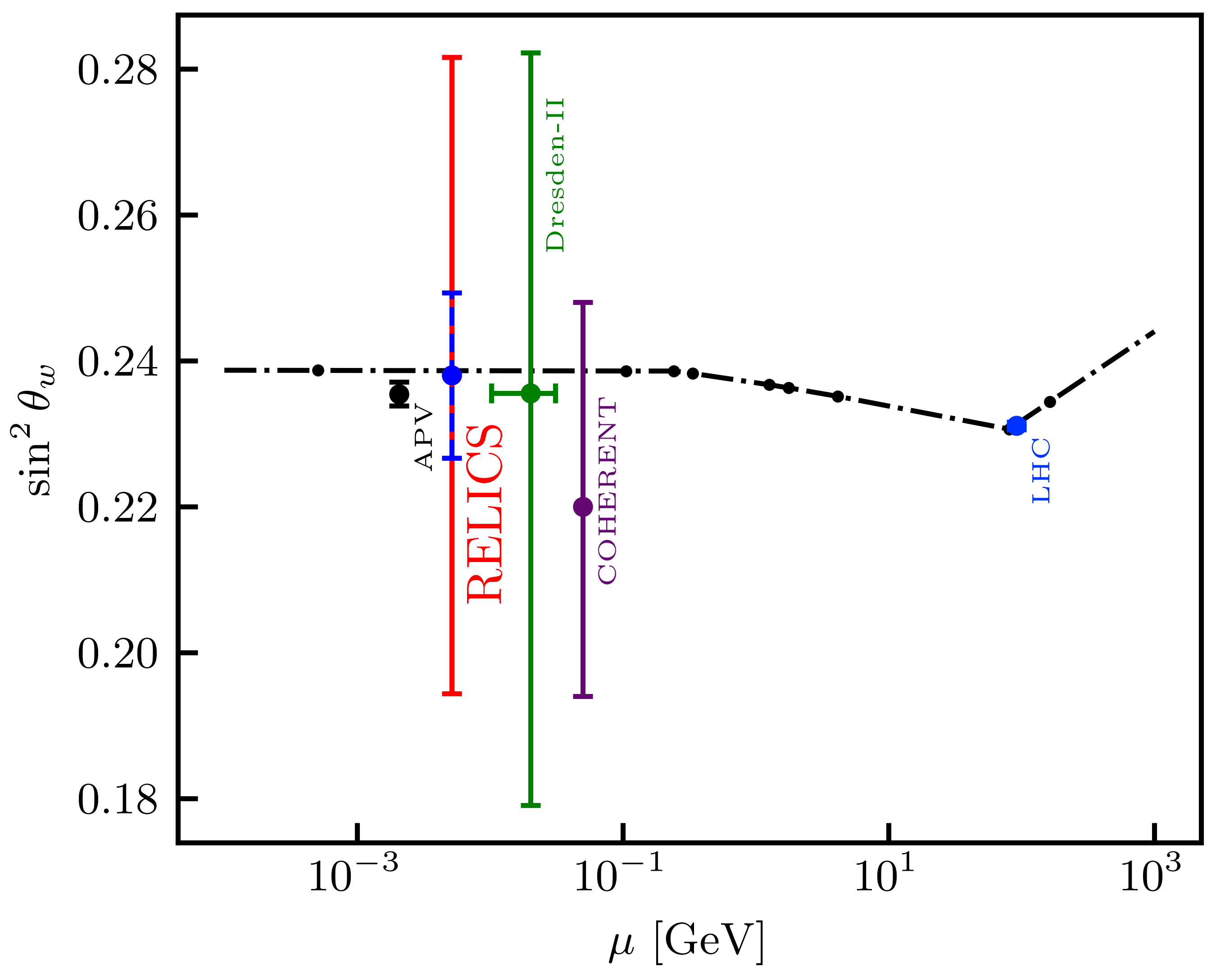}
\caption{The expected measurement (red) on $\sin^2{\theta_\mathrm{w}}$ with RELICS. The result without Q$_y$ uncertainty is overlaid with a blue-dashed error bar. 
The black line represents the dependence of the weak mixing angle in the $\overline{\rm MS}$ renormalization scheme\,\citep{Erler:2004in}. A comparison with experimental measurements from COHERENT~\cite{COHERENT:2021xmm}, Dresden-II~\cite{Majumdar:2022nby}, atomic parity violation (APV)~\cite{Guena:2004sq}, and LHC~\cite{ParticleDataGroup:2018ovx} are also shown.}
\label{fig:nu_weak_globle}
\end{figure}

The expected constraint on $\sin^2{\theta_\mathrm{w}}$ from RELICS is shown in figure\,\ref{fig:nu_weak}. The great improvement from the red solid line to the blue dashed line indicates that the charge yield variation is a major contribution to the constraint uncertainty. The sensitivity is expected to be competitive and improve the Dresden-II's result~\cite{Majumdar:2022nby} measured with reactor neutrinos. As illustrated in figure\,\ref{fig:nu_weak_globle}, RELICS will provide a measurement on $\sin^2{\theta_w}$ with momentum transfer down to the MeV scale, which is a valuable complement to the current measurements.

\subsection{Nonstandard neutrino interactions}
\label{subsec:nsi}
The detection of \cevns provides a new probe to study neutrino physics beyond the SM, such as the NSI of neutrinos. NSI has been extensively studied in the literature\,\citep{Barranco:2005yy}. This work focuses on a model-independent approach in the neutrino-quark sector. The parameterization of the NSI contributions to the cross-section can be expressed as Eq.\,\ref{eq:cevns_nsi}, which indicates that the cross-section strongly depends on the effective electron neutrino-up quark and -down quark interaction strength parameters of $\epsilon^{uV}_{ee}$ and $\epsilon^{dV}_{ee}$.

\begin{equation}
\begin{aligned}
\label{eq:cevns_nsi}
\footnotesize
\begin{split}
    \frac{d\sigma}{dE_{R}} = &\frac{G^2_F\cdot m_t}{2\pi}F^2 \left(2m_tE_{R} \right)\left[2-\frac{m_tE_{R}}{E_{\nu}^2} \right] \\
    & \times \left[Z(g_V^p + 2\epsilon_{ee}^{uV} + \epsilon_{ee}^{dV})+N(g_V^n + \epsilon_{ee}^{uV} + 2\epsilon_{ee}^{dV}) \right]^2
\end{split}
\end{aligned}
\end{equation}

\begin{figure}[htbp]
\centering 
\includegraphics[width=.45\textwidth]{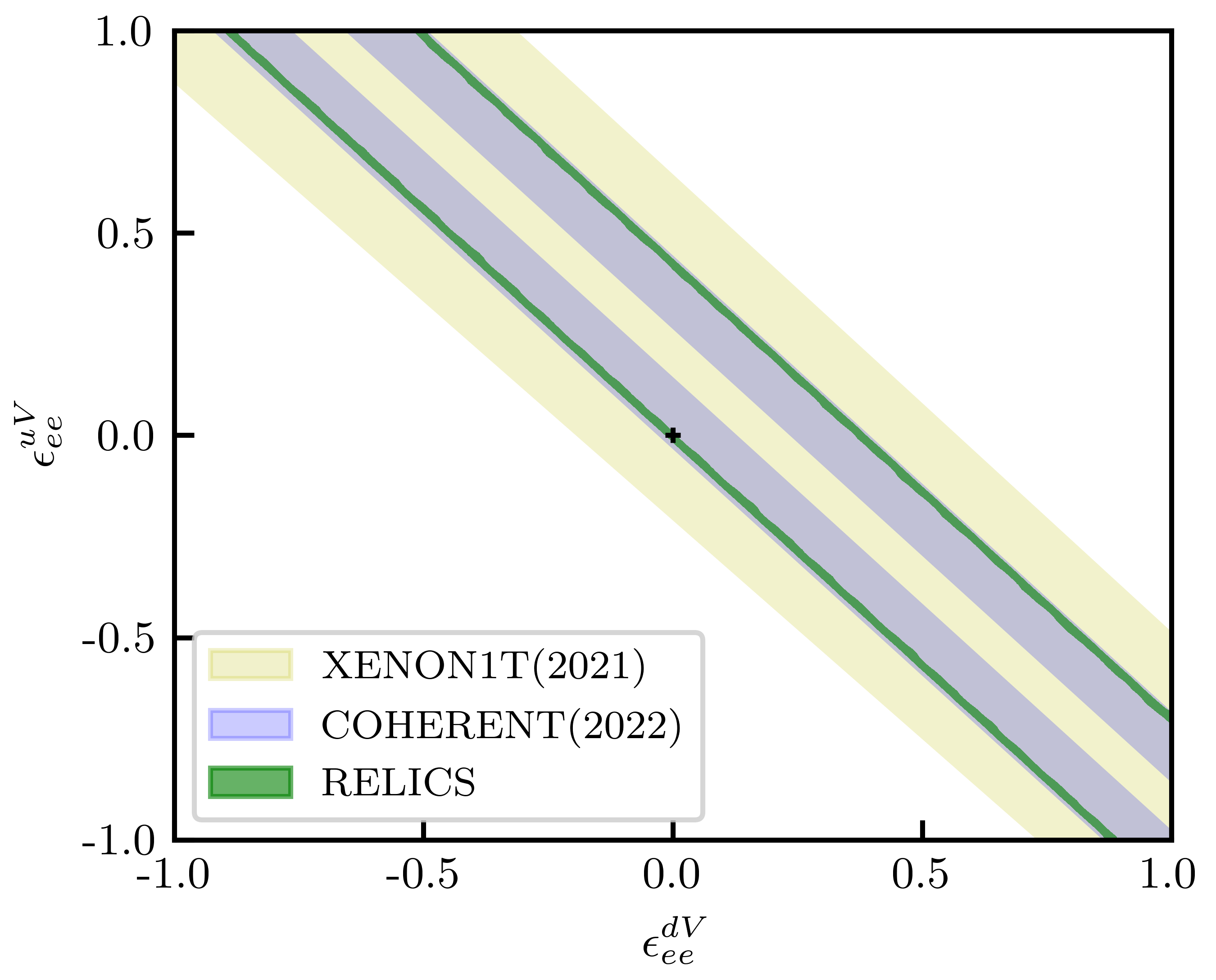}
\caption{The 90\% allowed parameter space of ($\epsilon^{uV}_{ee}$, $\epsilon^{dV}_{ee}$) constrained by RELICS detector. The COHERENT\,\citep{COHERENT:2021xmm} and XENON1T\,\citep{XENON:2020gfr} results are superimposed for comparison.}
\label{fig:nu_nsi}
\end{figure}

Eq.\,\ref{eq:nsi_s2_ll} defines the $\chi^2$ of likelihood, where all the parameters are the same as Eq.\,\ref{eq:thets_s2_ll}, except $\epsilon^{uV}_{ee}$ and $\epsilon^{dV}_{ee}$. All the related uncertainties of nuisance parameters are also taken in Tab.\,\ref{tab:thets_s2_uncert}. Figure~\ref{fig:nu_nsi} shows the constraining on $\epsilon^{uV}_{ee}$ and $\epsilon^{dV}_{ee}$ of RELICS under the standard model assumption, where both of the parameters are zero. The latest result from COHERENT\,\citep{COHERENT:2021xmm} is overlaid for comparison. A tighter constraint can be achieved for the RELICS experiment with higher statistics.

\begin{equation}
\begin{aligned}
\label{eq:nsi_s2_ll}
\scriptsize
\begin{split}
&\chi^2 = \\
&\sum_j\left[\frac{\left(N_{\mathrm{obs},j} - N_{\mathrm{exp},j}(\epsilon^{uV}_{ee}, \epsilon^{dV}_{ee}, t)(1 + \gamma_f) 
    - \sum_{i}^3 B_{i,j}(1 + \beta_i)\right)^2}{\sigma^2_{\mathrm{stat},j}}\right] \\
&+ t^2 + \sum_{i}^3(\frac{\beta_i}{\sigma^2_{\beta_i}})^2 + (\frac{\gamma_f}{\sigma^2_{\gamma_f}})^2
\end{split}
\end{aligned}
\end{equation}

\section{Summary}
\label{sec:summary}

The RELICS experiment is a $\mathcal{O}$(100)\,kg LXe detector with a fiducial mass of 32\,kg that aims to observe the process of reactor CE$\nu$NS. The experiment is planned to be deployed at the Sanmen Nuclear Power Plant ($\sim$3\,GW) in Zhejiang province, China, with a distance of $\sim$25\,m, an average neutrino flux of $\sim$10$^{13}$cm$^{-2}$s$^{-1}$.

In this work, a preliminary design of the RELICS experiment was presented. A dedicated MC simulation has been performed to study the background contributions. The dominant backgrounds come from the Pile-up DEs.
The second largest background, nuclear recoils induced by cosmic-ray neutrons, is suppressed via a passive water shield. An ionization-only analysis channel will be adopted for RELICS analysis to achieve a sufficiently low energy threshold for \cevns detection. Its sensitivity to weak mixing angle and neutrino NSI are explored with 32\,kg$\cdot$yr exposure. 

The RELICS experiment will provide rich opportunities to study Beyond the Standard Model (BSM) physics and potential applications in nuclear reactor monitoring for nuclear safeguards.

\vspace{20pt}
\textbf{Acknowledgements.} We acknowledge CNNC Sanmen Nuclear Power Company for hosting RELICS. We thank Dr. Xun Chen for supporting the BambooMC package and Prof. Kaixuan Ni for fruitful discussions. RELICS is supported by a National Key R\&D grant from the Ministry of Science and Technology of China (No. 2021YFA1601600), a grant from the Natural Science Foundation of China (No. 12275267) and the Fundamental Research Funds for the Central Universities, Sun Yat-sen University (No. 22lglj11). 
Yifei Zhao's work on RELICS is supported by a grant for undergraduate research from Beijing Natural Science Foundation (No. QY23088).

\bibliography{bibliography.bib}

\end{document}